# The Species Problem and Its Logic
## *Inescapable Ambiguity and Framework-relativity*

**Steven James Bartlett**

e-mail: *shartlet* [at] *willamette* [dot] *edu*

## ABSTRACT

For more than fifty years, taxonomists have proposed numerous alternative definitions of species while they searched for a unique, comprehensive, and persuasive definition. This monograph shows that these efforts have been unnecessary, and indeed have provably been a pursuit of a will o' the wisp because they have failed to recognize the theoretical impossibility of what they seek to accomplish. A clear and rigorous understanding of the logic underlying species definition leads both to a recognition of the inescapable ambiguity that affects the definition of species, and to a framework-relative approach to species definition that is logically compelling, i.e., cannot *not* be accepted without inconsistency. An appendix reflects upon the conclusions reached, applying them in an intellectually whimsical taxonomic thought experiment that conjectures the possibility of an emerging new human species.

———————————————————

# THE SPECIES PROBLEM AND ITS LOGIC
## *Inescapable Ambiguity and Framework-relativity*

## CONTENTS

# THE SPECIES PROBLEM AND ITS LOGIC
## *Inescapable Ambiguity and Framework-relativity*

**Steven James Bartlett**

*The species concept is one of the oldest and most fundamental in biology. And yet it is almost universally conceded that no satisfactory definition of what constitutes a species has ever been proposed.*

— Dobzhansky (1935, p. 344)

*Have enough words been said and written on the subject of what species are? How many evolutionary biologists sometimes wish that not one more word, in speech or text, be spent on explaining species? How many biologists feel that they have a pretty good understanding of what species are? Among those who do, how many could convince a large, diverse group of scientists that they are correct?*

*At this last and most essential task, many great scientists have tried and failed. Darwin, Mayr, Simpson and others have taught us about species, but none has been broadly convincing on the basic questions of what the word 'species' means or how we should identify species. For its entire brief history, the field of evolutionary biology has simply lacked a consensus on these two related questions.*

— Jody Hey (2001), p. 326.

## INTRODUCTION

In biology the concept of species has had a long and tattered history. It is an ancient and a fundamental concept that author after author has attempted to define in different ways, and yet never one of them with sufficient justification to persuade the scientific world of its unique and universal applicability.

We are inheritors of a tradition that goes back more than two thousand years when Aristotle (384–322 BCE) laid the basis for taxonomy by classifying organisms by kind and by giving each kind two names: a name for the genus to which it belongs and a secondary name for the property that distinguishes that kind. Although taxonomists continue to formulate such binomial definitions of species, no one has yet been able to formulate a satisfactory, unanimously accepted definition that expresses what the concept of species means or should mean.

There are, as we shall see, very good reasons that explain why this should be the case. We shall see why, on the one hand, from the standpoint of the logic underlying species definition, it cannot be otherwise. On the other hand, recognizing the constraints of logic for what they are, we'll find that it is possible to define species successfully in a way that is conceptually iron-clad, that is, cannot *not* be accepted without inconsistency.

The species problem has important, wider dimensions than are often perceived. The problem relates not only to the definition of species, but to fundamental problems in human and machine pattern recognition, and to some of the most abstract studies of the nature of theory. The species problem is in modern guise at one and the same time the medieval problem of universals and the problem of natural kinds, while its answer hinges on a resolution of the age-old conflict between realists and subjectivists. The species problem is not often understood from a sufficiently wide perspective that appreciates its place as an important problem with wide-ranging consequences.

In writing this monograph I recognize that it presents a critical challenge to a tradition spanning two millennia; it cannot help but scandalize and perhaps affront some readers. A short description of the basis for the discussion that follows may therefore be appropriate, providing reference to some of the author's publications that give further explanation and justification.

I hold the view that scientists and ordinary people alike make statements and adhere to ideas which, when analyzed from a strictly developed, theoretically self-critical, epistemological standpoint, can be shown to exceed their own capacities both to *refer* and to *know* what they are talking about. When people do this, as they commonly do, both in everyday life as well as in scientific theorizing, they *trespass* beyond the boundaries of possible sense. They think and make statements about alleged referents that cannot possibly be referred to—as they believe they are doing—by them or by anyone else. They are, in an epistemological sense, caught up in forms of *projective thinking* and in ways of expressing themselves that are fundamentally illusory and inconsistent with their own presuppositions.

The present discussion of the logic of species definition builds upon my previously published work which shows that both of the widely employed concepts of *objectivity* and *subjectivity* are inherently projective, and I therefore point to ways in which we can do without either. Similarly, I recognize that both notions of the *discovery* and *invention* of species cannot, in principle, be justified, because, once again, both force their users to trespass beyond the limits of possible reference. It will follow from this approach that species are neither discovered nor invented; what their status therefore is will become clear.

Last, I recognize that all claims to knowledge have meaning and potential truth, falsity, provability or impossibility of proof, only relative to appropriate frames of reference in terms of which those claims can be thought, expressed, confirmed, or disconfirmed. The theoretical position resulting from this perspective I have called *framework-relative epistemology*. Its central purpose is to increase awareness of the frames of reference that we rely upon when we think and express claims, and to insure our compliance with the inherent limitations of the systems of reference of which we make use. Often, these frameworks are employed implicitly and remain in the background until we make a reflective effort to render them explicit.[1]

---

[1] The justification for the claims in this and the preceding two paragraphs is found in Bartlett (1971; 1975; 1976; 1982; 1983; 2005, Part III; 2011, Chapters 2, 9).

# 1. The Species Problem

*No one definition has satisfied all naturalists; yet every naturalist knows vaguely what he means when he speaks of a species.*
— Darwin, *On the Origin of Species* (1859, Chap. 2)

*[I]t is a hopeless endeavour to decide this point on sound grounds, until some definition of the term 'species' is generally accepted, and the definition must not include an element that cannot possibly be ascertained....*
— Darwin, *The Descent of Man* (1871, Chap. VII)

By the time of Carl Linnaeus (1707-1778), a wide variety of alternative systems for classifying plants was in use; the same plant was often given different names and placed in differing systems of classification. In his 1735 *Systema Naturae*, Linnaeus presented an integrated classification system that could, he believed, organize all organisms, whether animals or plants, from the level of kingdoms down to species. It was an attractive, well-developed system of categories, but despite its later influence, taxonomy has labored unsuccessfully to find an effective and unitary classification system that could integrate, in a way satisfying to all biologists, the findings of the developing science.

Before Darwin, debates were already in progress over the question whether common traits or shared reproduction should be used as defining criteria to distinguish separate species (Grant 1994). Darwin was aware of the problem, as the quotations at the beginning of this section show, but he may not have advanced appreciably beyond a state of ambivalence in conceiving how the term 'species' should be defined. In 1856, for example, he wrote to botanist Joseph Hooker:

> It is really laughable to see what different ideas are prominent in various naturalists' minds, when they speak of 'species'; in some, resemblance is everything and descent of little weight—in some, resemblance seems to go for nothing, and Creation the reigning idea—in some, sterility an unfailing test, with others it is not worth a farthing. *It all comes, I believe, from trying to define the indefinable.* (F. Darwin 1887, Vol. 2, p. 88, italics added)

Then, three years later, in *On the Origin of Species*, he wrote, "I look at the term species as one *arbitrarily given for the sake of convenience to a set of individuals closely resembling each other*, and that it does not essentially differ from the term variety" (Darwin 1964/1859, 52, italics added; note that the context here is the possible distinction between 'variety' and 'species').

And then, some three hundred pages later in the same book, he wrote: "[f]rom the first dawn of life, all organic beings are found to resemble each other in descending degrees, so that they can be classed in groups under groups. *This classification is evidently not arbitrary* like the grouping of the stars in constellations" (Darwin 1964/1859, 411, italics added; here the context is whether taxonomic classes are real or imagined).

A modern reader may shake his or her head over what on the surface appears to be outright ambivalence or perhaps flip-flopping on Darwin's part: First there is his claim that the term 'species' is "indefinable," then his assertion that its definition is arbitrary and chosen as a matter of convenience, followed by an expression of belief that classifications are not arbitrary at all but correspond to real groupings. The contexts in which these different claims were made are certainly different, but in the end no definition of 'species' emerges. This is not the place to engage in Darwinian textual analysis; instead, such quotations may be taken as evident signs of the historical deep-rootedness of the species problem.[2] If these quoted passages from Darwin's writings point to an ambiguity in his mind, it was later literally to prove to be, as this monograph seeks to demonstrate, a telltale, significant, and, in fact, an inescapable kind of ambiguity.

The species problem has arisen to frustrate generations of taxonomists due to the multitude of often incompatible ways that have been proposed to classify organisms. The problem is that there so far appears to be no unique and adequately general way to define the meaning of 'species'; the multiplicity of distinct ways of classifying organisms reflects a corresponding multitude of distinct ways of identifying and recognizing properties that can be specified so as to group organisms in separable classes.

Perhaps the earliest researcher to recognize the close connection between species definitions and criteria of identification was Ernst Mayr (1942, 1982, 2000). As the species problem gained momentum, biologists have recommended a variety of alternative defining properties and criteria of identification based on them—for example, Mayr's influential biological species concept ("groups of actually or potentially interbreeding natural populations which are reproductively isolated from other such groups" (Mayr 1942, p. 120)), or the phylogenetic species concept (a species is identified as the smallest set of organisms that shares a common ancestor and can be distinguished from other such sets), and then there is the straightforward criterion of identification, that of shared reproduction. Other biologists have favored an organism's genetic blueprint and place in an evolutionary map of descent; still others have argued for the inclusion of an organism's behavioral traits; and this listing could be expanded to include a veritable multiplicity of alternative preferred defining properties of species that have been urged by their adherents. During only the past three decades, more than twenty different definitions of 'species' have been advocated by biologists (Mayden, 1997; Hey, 2006), while among philosophers of science discussions and debates concerning the problem continue unabated without any clear-cut resolution.

The species problem is not new, but clearly the proliferation of competing species definitions and the amount of contention and head-scratching this has led to have made the species problem particularly prominent in recent years, and in need of an actual solution that is at once definitive, provable, and conclusive.

---

[2] For detailed discussions of Darwin's views on the definition of 'species', see, e.g., Mayr (1982), Beatty (1985), Stamos (2007), Wilkins (2009).

# 2. What Definition Can and Cannot Accomplish

*Controversies from the careless employment of terms ought to be impossible, and they can be prevented mainly through the agency of Definition. Well-defined words, clearly understood and intelligently expressed meanings, are a sort of panacea for the thinker; and, in proportion as we approach the ideal here or recede from it, we may expect accuracy and progress in thought or deterioration and confusion.*

– Davidson (1885, p. 2)

*[N]o problems of knowledge are less settled than those of definition, and no subject is more in need of a fresh approach.*

– Abelson (1967, p. 314)

During taxonomy's long history, the definition of species has had several purposes: Foremost among these is effective communication among biologists, so that, as a group, they may achieve a measure of consensus and consequent efficiency in organizing and communicating about the wide range of organisms. In serving this goal, however, a long tradition has it that species definitions are not believed to be mere conceptual constructions that have no anchor in reality, but instead are intended and are believed to refer to *real*, shared commonalities among distinguishable classes of organisms—commonalities that, as a result of their mind-independent reality, are *discovered*. Another long tradition has disagreed, claiming that, on the contrary, species definitions are constructs or artifices that human beings create; the commonalities they identify do not exist in reality, but only in people's minds; species are therefore *invented*. Independently of their opposing beliefs, both parties accept that species definitions are to be formulated in an authoritative context with the expectation that definitions judged to offer a true and comprehensive account of the evidence will persuade others to accept them. A proposed definition of species is intended to account for the facts, establish authority, and to persuade its acceptance.

There are different kinds of definitions, of which a few should be clearly distinguished here.[3] There are *stipulative definitions* that, through allegedly authoritative assertion, specify how important terms are to be understood. In addition, there are *real definitions* that embody empirical content; they are intended to refer beyond the mere wording of a definition to real referents that make the definition true or false. For example, a definition of DNA commonly includes specific information or explanatory content: "DNA is composed of the following nucleic acids ... that serve as the molecular basis of heredity, are localized in cell nuclei, are constructed in the form of a double helix, (etc.)....". A third variety of definition relevant to the task of defining species makes up the class of *coordinative definitions*, which assert relationships between the terms of a theory and the phenomena the theory is about.

---

[3] For a more comprehensive and detailed inventory and discussion of general types of definition, without application to the species problem, see Bartlett (2011, Chapter 2) and the earlier work, Robinson (1962/1950).

These three varieties of definition are routinely employed by theorists who propound opposing views concerning the *ontological status* of what they define—specifically whether what is defined exists in a mind-dependent or mind-independent way. These views, for which, as we shall see, justification cannot in principle be provided, I will call ontological "*biases*" since they rate no higher estimation.

During the past millennia, a deep divide has come to separate two contesting camps, each of which consists of more specialized sub-groups, whose individual agendas will not concern us here: They include realism, objectivism, and Platonism in mathematics, which stand together on one side of the ontological abyss, and together assert autonomous existence that is independent of human minds, while idealism, subjectivism, conceptualism, and intuitionism in mathematics stand on the other side, together asserting mind-dependence. Within a wide range of disciplines, these different –isms express opposing views that have come to separate theorists with respect to their ontological preferences.

The ontological conflict between these opposing groups is not merely an invention of the philosophers. It permeates human consciousness, whether in religion, psychology, mathematics, philosophy, or theoretical physics—wherever there can be disagreement whether the contents of awareness or theory or dogma exist independently apart from that awareness, theory, or dogma—as realists, objectivists, and Platonists claim—or whether those contents fail to possess an autonomous existence of their own—as idealists, conceptualists, subjectivists, and intuitionists believe. In what follows, for simplicity I will refer to the two opposing camps merely as "realists" and "subjectivists."

Species definitions as they are proposed by realists combine all three of the above roles of definition: They are at once stipulative; they are intended to have referents believed to possess a mind-independent existence about which such definitions make true statements; and they serve as coordinative definitions that establish relationships between the systematic nomenclature of a taxonomy and specified classes of organisms. Species definitions viewed as human constructs also are stipulative and serve as coordinative definitions, but deny they have a basis in mind-independent reality and therefore assert they are only human constructions.

As we've noted, the core of the species problem for any taxonomist who proposes a species definition lies in his or her wish to provide a conception of species that serves as a unique, univocal, comprehensive, and persuasive definition. However, in addition, and consistently overlooked in the literature, the species problem is more deeply problematic because of inconsistent thinking that is internal both to the realist claim and to the subjectivist claim. As we shall see, those taxonomists who assert that their definitions have real, mind-independent reference make one kind of conceptual mistake, while taxonomists who claim that species definitions are mere human constructs make another. Compounding these errors, we'll also find that the very wish that both camps express—to formulate one univocal and comprehensive species definition—is, for logical reasons alone, impossible.

To take things one at a time, we'll first examine the framework presupposed by realist species definitions—believed to have real, objective, mind-independent reference. Later (in §5), we'll examine the view of the subjectivists who claim that the commonalities their definitions recognize have the status of fictions, are "only subjective," and hence are "invented."

The definition of 'centaur' is not a real definition, neither are the definitions of 'line' and 'point' in Euclidean geometry. Real definitions make assertions, believed to be true, of objects asserted to have mind-independent reality. It is here that real definitions, and hence definitions of species that are presented as real definitions, get into trouble.

Human psychology of course plays a role. Mathematician and philosopher Alfred North Whitehead (1925) described "the fallacy of misplaced concreteness," the mistaken reifying belief that what is a fiction or an abstract construct exists physically. The mistake is psychologically easy for many people to make, and this is most especially a self-destructive deterrent to progress when it happens in science, as we recall from Ptolemy's epicycles, phlogiston, the luminiferous aether, bodily humors, the vital force, Darwin's gemmules, and extending perhaps to today's inferential-hypothesized realities that include dark matter in cosmology and branes in string theory.

Reification is not just a technical misstep of faulty logic, but it is basic to the human psychological constitution. The readiness to believe that, once defined, the thing defined should exist in its own right, is a predisposition of human psychology. I have studied this propensity elsewhere (Bartlett 1971; 1975a; 1976; 1982; 1983; 2005, Part III; 2011, Chapters 2, 9); relevant here is the psychological fact that, once we formulate definitions that we mistake to be real definitions, we build on these, using them to make further statements that we then take to be true of their purported referents. By then, the ties that should bind us to our original framework of reference have been sufficiently loosened, and the ties that should anchor our thoughts and their expression tend to be forgotten, so that we no longer are aware of what it is we are actually referring to and talking about, all the while believing that what we have in view possesses an independent reality of its own.

I have called this psychological propensity to believe in alleged realities that are, so to speak, "thrown beyond the boundaries of our systems of reference," *epistemological projection*. In the context of the logic of species definition, we fall victims to our own propensity to engage in epistemological projection when we are persuaded to believe that stipulated, coordinative definitions are actually real definitions that refer to independent empirical realities.

When realist species definitions are propounded, a psychological and illogical shift often takes place: Having stipulated a certain species definition, that definition is transformed into a judgment that there exist, in the organisms being classified, shared real commonalities of properties that at once define the class to which they belong, and distinguish that class from others. We shall see how this shift that *defines species into existence* is not only a fundamental mistake of epistemology, but runs headlong into an inescapable limitation of judgments of commonality.

## 3. Classification and Pattern Recognition

*Toutes les activités intellectuelles humaines commencent par la formation mentale de classes d'objets. Cela est vrai non seulement dans les sciences*

> *classificatrices et dans les sciences physiques mais aussi dans la vie quotidienne aussi bien que dans les problèmes technologiques et médicaux.*
> — Watanabe (1965, p. 39)
>
> All human intellectual activities begin with the mental formation of classes of objects. This is true not only in the classificatory and in the physical sciences, but also in daily life as well as in [the study of] technological and medical problems.[4]

As Japanese physicist Satosi Watanabe (1910–1993) has observed, classification is conceptually fundamental in science as well as everyday life. Without classification we would be unable to make abstract, organized sense of the world; we would be limited to the undifferentiated experience of non-reflective, non-intellectual present sensation, memory of past impressions, and unstructured apprehensions of the future. We could call on no integrating concepts and hence could formulate and communicate no general principles by means of which to establish an order and structure based on experience: Generalization, deduction, and induction would all cease to be possible.

The mental ability to identify classes of objects is indeed basic, and, as logicians would say, it is a logically primitive (i.e., fundamental) capacity without which intelligent, meaningful, and useful comprehension becomes impossible.

Classification ultimately rests on the *pattern recognition of commonalities*, the ability to identify shared properties, or features, or structures. To understand the nature and limits of classification, we need to understand the logic that governs the recognition of commonalities, to which we shall turn in the next section.

The study of pattern recognition, and in particular the modern study of what I will here call the *logic of commonality*, together have emerged from a long tradition that has focused on the nature and ontological status of so-called *universals*—of general ideas, names of properties, universal truths, etc. Plato and Aristotle took positions on this question, as did their intellectual descendents a thousand years later, including William of Champeaux, Abelard, Scotus, Aquinas, and Occam, and, yet another thousand years later, as is to be expected of perennial philosophy, the question persists today in the still-unsettled controversies among realists and subjectivists (as well as the more ontologically neutral group of nominalists, who believe that universals or properties are simply words, with no real or mental referents).

More evident progress in the study of pattern recognition has been made in computer science and in psychology and cognitive neuroscience.[5] In computer science, studies of machine pattern recognition have led to substantive and useful results. Computerized pattern recognition has now become woven into our daily lives—examples include speech recognition; optical character recognition; machine recognition of handwriting; face detection by cameras; and fingerprint, license plate, and facial recognition by police computers.

---

[4] This and subsequent translations from the French are the author's.

[5] Relating to machine pattern recognition, see, e.g., Watanabe (1985), Theodoridis & Koutroumbas (2009/2006), Rokach (2010), Kuncheva (2014); in connection with the psychology of pattern recognition, see, e.g., Watanabe (1985), Margolis (1996), Eysenck (2001).

Some of these advances, now highly technical and mathematicized, have profited from research findings in psychology and cognitive neuroscience where human pattern recognition has come to be understood as a matching process that compares present perceptions with remembered information. This human-centered work has led to several leading theories of pattern recognition, among them template and prototype matching: According to the first theory, sensory data are compared with pattern models, or templates, that are stored in memory (for example, the alleged mental "template" that enables one to identify the following as the same number: **5**, *5*, 5, 5). In the second theory, the properties of a given subject of reference are thought to be "averaged" in a concept that serves as a prototype for future identifications (for example, the concept of a cold-blooded marine vertebrate with fins can serve as a prototype of a shark, swordfish, grouper, barracuda, tuna, etc.)

Despite such advances, research both in psychology/neuroscience and in computer science has often fallen victim to the projective variety of conceptual error referred to earlier, while philosophical thought is no exception. In all three of these areas we see evidence of the human urge to project, to transgress beyond the limits of possible reference. In philosophy, the subjectivists, nominalists, and realists mentioned above have argued for millennia whether universals exist only in the mind, only in language, or in combination, or whether they exist independently of minds and language. In philosophy of science, a long history of debate similarly continues to be fought in efforts to establish whether natural kinds—the grouping of natural objects—once again reflect the discovered structure of an independently existing natural world, or only psychologically-based human constructs that are invented. Later, we'll explore why such epistemological projections are fundamentally self-undermining and without meaning, and will look at an instructive case of projective thinking in one attempted approach through computer science to understand pattern recognition.

Psychologists have attempted to distinguish real from imagined patterns. As one would expect, real patterns are thought to be patterns that putatively exist "in reality," whereas imagined patterns are patterns that "don't actually exist." In this view, when you see that **5**, *5*, 5, 5 have certain properties in common, you have recognized a pattern that is said to be real. However, when you look at the full moon and recognize the image of a man's face, you don't see what is real—so it is claimed; the pattern that you see, according to this view, is only imagined. Psychologists have named the human propensity to imagine patterns *apophenia*,[6] and have gone further to name the (alleged) misperception of random data *pareidolia*.

---

[6] The term (*Apophänie* in German) was coined by Klaus Conrad in his 1958 study, *Die beginnende Schizophrenie: Versuch einer Gestaltanalyse des Wahns* (The Onset of Schizophrenia: Attempt toward a Gestalt Analysis of Delusion). Conrad created the word '*Apophänie*' to refer to the onset of delusional thinking in psychosis. 'Apophania' has come to mean the imagined (or delusional) seeing of patterns—that is, patterns "where none really exist." The word is commonly applied, for example, to a gambler's tendency to see patterns in game play

Apophenia and pareidolia are supposed to be, in other words, fundamentally different from the perception of "real" patterns. To mark this contrast, we may say that, according to this conventional account, the two types of patterns reflect different *ontological biases*. Since the majority of people recognize that **5**, *5*, 𝟝, 5 are symbols for the same number, an ontological bias in favor of realism is established by general consensus; whereas if you see a face in a cloud shape, and most other people don't, there is little bias in favor of realism. Recognizing the Man in the Moon, however, is something that a great many people see, and here the boundaries between real patterns, apophenia, and pareidolia become indistinct and arbitrary.

To make headway in connection with the species problem, a clear and precise understanding of the recognition and identification of patterns will require something more than sloppy thinking of this kind. In particular, we shall need a better grasp and formulation of the logic of commonality.

## 4. The Logic of Commonality

> *Modern logic ... has the effect of enlarging our abstract imagination, and providing an infinite number of possible hypotheses to be applied in the analysis of any complex fact.... It has, in my opinion, introduced the same kind of advance ... as Galileo introduced into physics, making it possible at last to see what kinds of problems may be capable of solution, and what kinds must be abandoned as beyond human powers. And where a solution appears possible, the new logic provides a method which enables us to obtain results that do not merely embody personal idiosyncrasies, but must command the assent of all who are competent to form an opinion.*
> 
> — Russell (1972/1922, pp. 68-69)

Despite widespread progress in science, a shade of suspicion is often felt by researchers as well as by the general public when highly abstract mathematical results are applied to their specialties and to everyday reality. In connection with our subject-matter here, this has been particularly true of those biologists who continue to advance and debate competing definitions of species. What we have learned from purely abstract and formal studies provides a technical framework that dissolves such disagreements, and points to a satisfactory and, as we shall see, a theoretically compelling answer to the species problem. In publications relating to the species problem, the mathematical results that are summarized in this monograph have been ignored; they are long overdue a careful hearing and emphasis.

A personal anecdote from the author's experience may illustrate the reluctance of many researchers to embrace the applicability of purely formal results to their more concrete areas of interest. In the early 1970s, I served as a consultant to the Rand-National Science Foundation Project in Regional Analysis and Management

---

"that don't exist," or to the recognition by fervently religious people of the face of a saint in the graining of a piece of wood or in the shape of a Cheeto®.

of Environmental Systems (Bartlett, 1975b).[7] The main goal of the project was to develop machine algorithms capable of detecting previously unspecified environmental commonalities from among unstructured data. The intention, in other words, was to develop a variety of computer-based pattern recognition that could be applied to environmental management, with the specific objective of developing computer algorithms capable of sorting unstructured data so that commonalities among the data would be disclosed and made explicit to the human operators.

This goal of course antedated the substantial developments that were to come later in the field of pattern recognition, and so the project was characterized by a certain amount of understandable innocence. The conceptual naïveté that concerned me as a consultant was the prevailing unquestioned belief among the project's participants that it should be possible to detect commonalities from among disorganized data *without* first specifying the criteria by virtue of which those commonalities could be identified. Clearly, if such commonalities were to be specified in advance, a computer algorithm might be developed to detect them—but it was the basic purpose of the project to design machine pattern recognition that could detect patterns of commonality *not specified ahead of time*. It was the belief of most of its participants that this could be done.

The reader should quickly see the connection between this belief and the similar belief among many biologists who have sought to formulate species definitions that identify taxonomic commonalities among groups of organisms. Much of the controversy in connection with the species problem is an expression of the fundamental and persistent belief that organisms do fall into classes by virtue of their "naturally occurring" commonalities—that is, commonalities that "are just there, that mark real distinctions that must be recognized in any valid, scientifically compelling system of classification."

At the time of the Rand-NSF project, I was familiar with some of the then-recent work of theoretical physicist Satosi Watanabe, which I described to the Rand-NSF group members. If Watanabe's unexpected and far-reaching proof were to be accepted by the project participants, then they would need to accept that it would be impossible to achieve the basic goal of the environmental management project. With characteristic American optimism, the bold rejoinder of one project participant was, "Well, we'll do it anyway!"

Unfortunately, it doesn't pay to ignore impossibility proofs: There is no point investing large sums of money and human effort in building a structure whose design exceeds the stress limits not only of any known material, but of *any possible organization of physical matter*. The logic of commonality provides this kind of impossibility proof.

## 4.1 *Similarity and Commonality*

The most theoretically fundamental studies related to our subject have focused attention on what has become known as *similarity theory*. Similarity theory attempts to understand and formulate principles that underlie and govern the possible detection of similarities, whether by human beings or by artificial intelligence. This previous work has concentrated attention on the concept of similarity rather than that of commonality.

---

[7] Headquartered at Colorado State University, under Grant GI-33370A#1.

There are, however, good reasons to consider commonality to be the more fundamental concept. Like other similarity theorists, Watanabe gave the concept of similarity center stage, and yet, even so, he recognized: "To persuade that two objects are similar, it is natural to enumerate the properties that are *commonly owned* by the objects. The more properties are *shared*, the more similar they are" (Watanabe 1985, p. 75, italics added). —In other words, we make recourse to detection of commonalities, the sharing of properties, *in order to judge* that objects are similar.

To explain this further, similarities between objects are identified as a function of their shared possession of a property—i.e., its commonality—which they may have to a variable degree. The property they may have in common is, in turn, itself identified as a function of certain, often fuzzily defined, limits. For example, in order to judge whether two objects are similar in having a particular hue of blue, one must have in mind some approximate standard of that color; it may be a vaguely defined standard, but even though fuzzy it can nonetheless be usable. We may not know, or agree precisely, where the boundaries of that hue of blue begin and end, but we usually are not for that reason prevented from accomplishing practical objectives.

Commonality differs from similarity in the following sense: When we recognize that two objects share the same property, we thereby detect a commonality between the two objects. This recognition may not require a human being to have a particular standard in mind, as is the case with similarity judgments, which require one to specify *with respect to what standard* recognition of similarity is made. Rather, the mere awareness that there is a commonly shared property, whatever it may be, can be enough; there may be no need to apply a separately held criterion.

In short, recognition of similarities requires that we make recourse to some standard, separate from the objects being compared and held independently in mind, relative to which property matching and thereby judgments of similarity can be made. Recognition of commonalities in some instances can be less dependent upon comparison relative to a separately conceived standard, as when people intuitively recognize certain commonalities between two objects without having been instructed what to look for. Such an intuitive form of recognition of commonalities seems, as we'll see, to be an inherent, even innate (but likely not exclusively human) natural ability.

The difference in the levels of logical primacy and of relative simplicity of the concepts of similarity and commonality becomes quickly evident in computer programming: It can be a great deal easier to program a computer to identify commonalities among objects, defined in terms of a certain specified property, than it is to write a program to detect similarities. For example, it not difficult to get a machine to identify when a specified facial feature—say, brown eyes—is shared by any two human subjects. But it is much more difficult to instruct a machine to recognize the similarity among faces that is designated by the expression 'baby-faced'. Once we have clearly defined 'brown eyes', it is a simple matter to detect them; but 'baby-faced' is essentially a similarity-identifying expression: Individual people can be baby-faced in a wide variety of overlapping ways; there is no one specific, unique property they must share. Recognizing when two people have brown eyes in common is a much simpler task. Commonality detection tends to lend itself much more to precise specification than does the detection of similarities.

For such reasons, among others there is no relevant need or space to discuss here, commonality rather than similarity will in the following discussion be regarded as the more fundamental concept.

Whether one prefers to use the term 'similarity' or 'commonality', we should note that both forms of identification are based upon the prior ability to identify specific properties. Being able to explain what it is that makes a certain judgment of commonality true rests on the more fundamental ability to identify when two or more objects each has the property in question.

We therefore have a simple hierarchical ranking of pattern recognition concepts and corresponding abilities, with the ability to detect a specific property on a lower, more fundamental level; on the next higher level of complexity, the ability to recognize when a given property is shared in common by two or more objects; and, on a still higher level, the ability to recognize similarities with respect to specified properties.

The ability to identify, recognize, name or otherwise refer to properties will here be left undefined as a logically primitive, unexplained concept. Although this is an important assumption to make, its legitimate status as an assumption is borne out by recent studies. To date, research in both the psychology of pattern recognition and the theory of machine pattern recognition point to the same tentative conclusion, that human subjects are "innately" predisposed in their perceptual experience to pick out properties that are, in a multitude of possible ways, important to them. In other words, they are selective with respect to those properties that attract their attention, what they then pay attention to, what they remember, and what they tend to anticipate. Even newborn infants display this attentional selectivity. When it is claimed that this ability is "innate," this is another way of saying that we don't ultimately understand why human beings are like this or how they manage to do this; for now, we simply accept that this is the way they are: The ability to detect properties is here then left unexplained as inborn, inherent, or instinctive.

Our present understanding of human pattern recognition therefore accepts that this ability to home in on an object's properties, properties that are perceived to be important in some way, is both psychologically and theoretically fundamental. Derivatively, the conceptual framework of pattern recognition by means of artificial intelligence, which seeks to simulate human abilities, also accepts that attentional selectivity plays an elemental role in the recognition of patterns. But machine pattern recognition is different from the human in that it is necessary to specify in advance what the machine is to look for.

When researchers claim that the human selective predisposition is "innate," this, on an abstract theoretical level, plays a role kindred to the status as an assumption that we give here to the identification, recognition, and naming of properties. From a logical point of view, a primitive concept, one that is assumed without inference from others, is, semi-metaphorically, "innate" to that logical point of view.

Even though we choose for present purposes to leave the ability to detect properties unexplained, it is important that we do not unknowingly import and leave unexamined another related and conceptually fundamental concept. That concept, which rides along parasitically, so to speak, on the back of property detection, is *valuation*. As soon as attention, human or machine, is selectively directed to some particular property, this implies some degree of recognition, implicit or explicit, of the *relative importance*, and hence *value*, of that property; for that property has to possess some degree of relative importance in order to be recognized as an object of attention discriminated from among an indefinite plurality of others that are possible. This "value-dimension" of property identification will become central to our later discussion.

## 4.2 *Similarity Theory and the Species Problem – Historical Landmarks*

There have been few significant landmarks in the development of similarity theory as it relates specifically to the species problem. This is largely due to the fact that discussions of the species problem by biologists have seldom taken into account studies of similarity theory, which have evolved on an abstract, formal, and often heavily mathematical plane. It is understandable that connections between the two independently evolving areas would not often be recognized.

An early paper more than half a century ago by H. Gulliksen (1947) recommended the usefulness of studying *paired comparisons* in connection with what he termed "the logic of measurement." Gulliksen's interest was directed toward the ways in which human subjects in psychological experiments made paired comparisons between adjacent members of an ordered series of test objects (for example, test subjects might be asked to make comparisons among different pairs of lights, each of which has a different intensity of brightness). Gulliksen did not relate paired comparisons to comparative judgments of similarity, but his work helped point to the efficiency of reducing comparative judgments to comparisons of objects taken a pair at a time. We shall see how this pair-wise strategy becomes important in the development of similarity theory.

It was not until the early 1960s that Satosi Watanabe, and then later Nelson Goodman, published research that made similarity theory directly relevant to the biological species problem. (Although neither Watanabe nor Goodman discussed the species problem *per se*, Watanabe was aware (as likely also was Goodman) that similarity theory applied in important ways to taxonomy).

Watanabe was a theoretical physicist with broad research interests that wove together information theory, mathematical logic, set theory, Boolean algebra, and lattice theory. As a young man he worked under the direction of Heisenberg, developing a counter-conventional and still not widely known study that sought to demonstrate that quantum probability is time-asymmetric, that is, that the probability laws describing quantum events are not reversible in time.[8] In 1962, Watanabe participated in an international conference on information theory and prediction in science held in Brussels, and there presented work with a direct bearing on the species problem, work seldom known by biologists and therefore not often recognized for its fundamental importance to taxonomy. This work will be summarized in the next section.

---

[8] For a relatively recent affirmation of Watanabe's results, see Holster (2003).

Nelson Goodman (1906–1998) was an American philosopher one of whose main interests was the logical study of parts and the wholes they form, which in mathematical logic has come to be known as mereology. In 1972, Goodman published a paper directly relevant to the species problem, and which, unlike Watanabe's more technical earlier contribution to the theory of taxonomy, is occasionally cited by biologists. In his 1972 paper, Goodman did not seem to be aware of Watanabe's work even though Goodman was writing a decade later. Goodman's paper, whose informally reached conclusions supported Watanabe's results, ought properly then to be considered subordinate to Watanabe's. Goodman's thoughts will also be summarized in the next section.

## 4.3 *Watanabe's Theorem of the Ugly Duckling: A Neglected Variety of Formal Limitation*

> *This paper contains a proof of a theorem which I should like to dub "Theorem of the Ugly Duckling," for it claims that an ugly duckling and a swan are just as "similar" to each other as are two swans. This would make H. Chr. Andersen's story pointless and all scientific efforts groundless. The aim of the paper is to save our beloved Andersen and science, showing that the actual process of classification of objects is based on an extra-logical factor beyond the grasp of such purely formal consideration.*
>
> – Watanabe (1965, p. 39)

The above abstract in English accompanied Satosi Watanabe's first published proof, written in French, of his "Theorem of the Ugly Duckling."[9] Because of the theorem's direct bearing on the species problem and the fact that the majority of biologists who have written about the problem have been unaware of Watanabe's work, a brief summary and worked out illustration of his theorem are given here.

Watanabe recognized the importance of his theorem and repeatedly presented it in his later publications, but he gave the fullest account in its earliest publication (Watanabe 1965). As is the case with his work relating to the irreversibility of time in quantum theory, Watanabe's theorem pushes beyond the limits of conventionally accepted views, and it comes as a surprise to many people.

---

[9] He first described the theorem in a paper given in Denver at the annual meeting of the American Association for the Advancement of Science in 1961. He then developed a formal proof of the theorem, which he presented in 1962 at the International Symposium on Information Theory and Prediction in Science, sponsored by the Académie Internationale de Philosophie des Sciences in Brussels. His proof was not published until 1965, and when it was, it appeared in French with the short English abstract quoted at the beginning of this section. The French text, however, makes no mention of ugly ducklings and swans; instead, it uses a different comparative analogy, between sparrows and crows. This suggests that the idea of naming his theorem after the ugly duckling of Andersen's story came to Watanabe as a happy after-thought that he then incorporated into the English abstract placed before the French text.

The Theorem of the Ugly Duckling reflects Watanabe's recognition that categorizing things in classes is fundamental to virtually all conceptual processes. I quoted part of the following passage earlier in this essay; here is the complete passage:

> *Toutes les activités intellectuelles humaines commencent par la formation mentale de classes d'objets. Cela est vrai non seulement dans les sciences classificatrices et dans les sciences physiques mais aussi dans la vie quotidienne aussi bien que dans les problèmes technologiques et médicaux. Le but de ce mémoire est d'une part de donner une formulation mathématique du processus de classement en précisant en un langage logico-mathématique ce que nous faisons souvent intuitivement, et d'autre part de donner un algorithme de classement qui peut être exécuté par une machine à calculer.*
>
> – Watanabe (1965, p. 39)

> All human intellectual activities begin with the mental formation of classes of objects. This is true not only in the classificatory and in the physical sciences, but also in daily life as well as in [the study of] technological and medical problems. The purpose of this paper is, on the one hand, to give a mathematical formulation of the process of classification by stating in logical-mathematical language what we frequently accomplish intuitively, and, on the other hand, to present an algorithm of classification that can be executed by a computer.

Among the most fundamental questions to be asked about classification, Watanabe claims, are these:

> *Existe-t-il une possibilité de classifier d'une façon unique un ensemble d'objets dont les propriétés nous sont connues? Est-ce qu'il y a une classification « naturelle » d'objets? Nous crayons naïvement que deux corbeaux partagent beaucoup plus de prédicats en commun qu'un corbeau et un moineau n'en partagent en commun. Notre analyse ... cependant montrera que cette croyance est entièrement fausse.* (Watanabe 1965, p. 40)

> Does any unique method exist for classifying a set of objects for which the properties are known? Is there a "natural" classification of objects? We naïvely believe that two crows share many more predicates in common than a crow and a sparrow share in common.[10] Our analysis ... however will show that this belief is entirely false.

---

[10] Why a crow and a sparrow here, and not a swan and a duckling? See footnote 9.

When confronted with unstructured information—data that has not already been selectively perceived and conceptualized—the general human tendency is to believe—once we recognize an order in that information, once we perceive a structure among the data—that the order we recognize and the structure we see "were there all along." This, reappearing once again, is an *ontological bias*, a belief that is questionable and which we will proceed later to do.

When we perceive two objects and recognize that they belong to the same class because they share certain specified properties, the general tendency is to believe that they share more properties in common than do any pair of randomly chosen objects. This is a *classification belief*, one which Watanabe's theorem leads one rationally to question. The classification belief is usually fused at the hip with the ontological belief. The resulting combined belief routinely takes the following form: When we perceive that two objects have more properties in common that other pairs of objects chosen at random, we believe ourselves, first, not to be victims of arbitrary judgment, and, second, we hold firmly to the additional belief that what we perceive is an independently existing reality. We need to back away from both of these uncritically held beliefs and acquire some dispassionate distance.

When confronted by two objects, one of which is a live swan and the other a live duckling, we have already advanced beyond a level of unstructured information and have selectively identified the objects as objects of particular kinds for which we apply different names. Yet, the set of the properties of those two objects, properties which are in principle identifiable, contains indefinitely numerous possible shared properties: Both objects share the property of being birds, of being perceived at a certain time and by a certain observer; both are, for example, less than a thousand years old (and of course they also have the indefinitely numerous additional properties of being less than a great many other ages!); both birds possess the property of being unable to speak English (and also the properties of not speaking any other human language from among the host of possible languages); both have the property of not being florescent in color, not being radioactive, not being highly charged electrically, and the list could go on and on—indefinitely; and so the list of the two birds' possible shared properties is potentially infinite.

Suppose we take the opposite approach, and list the properties the swan and the duckling do not have in common? One is large, the other small; one is an older animal, the other is very young; one weighs $x$, the other $y$; one is self-sufficient, the other not; one is white, the other grey; one is $n$ inches high, the other $m$ inches; one was born on a weekday, the other on a weekend; one has walked a total distance of $d_1$, the other $d_2$; etc. —In short, the properties they don't share are also indefinitely numerous.

Reasoning in this informal, impressionist way suggests that perhaps the number of properties the two birds have in common may not, in relation to the range of all the other properties they do and don't have in common, be enough to justify, from a strictly logical point of view, calling them both members of the same class. This impressionist reasoning points in the direction of Watanabe's theorem, but it has little theoretical compelling force. Watanabe sought to prove that the classificatory belief *must, of logical necessity, be mistaken*; later on, we'll show how the linked ontological belief is, as well.

Watanabe's proof relies on familiarity with mathematical logic, set theory,

Boolean algebra, and their combined application in a form of mathematical analysis that he called spectral decomposition. His *modus operandi* throughout his published work was highly technical. Certainly his heavily mathematicized style of presentation may have been responsible for the lack of attention his Theorem of the Ugly Duckling has received in connection with the species problem.

It is possible to gain a reasonable understanding and appreciation of both the theorem and its proof in a simpler and more effectively communicated way. Watanabe gives several clear statements of the consequences of his theorem. Here are several of them. The theorem[11] proves:

- "that it is impossible to speak of the similarity or dissimilarity of objects as long as one attempts to define these concepts as a function of the number of all the possible predicates that the objects possess or do not possess" (Watanabe 1965, p. 49).[12]
- "from the formal point of view there exists no such thing as a class of similar objects in the world, insofar as all predicates (of the same dimension) have the same importance. Conversely, if we acknowledge the empirical existence of classes of similar objects, it means that we are attaching non-uniform importance to various predicates, and that this weighting has an extra-logical origin" (Watanabe 1969, p. 376).
- "... any two objects are equally as similar to each other as any other two objects, and are equally as dissimilar to each other as any other pair, insofar as the number of shared predicates is regarded as a measure of similarity and the number of predicates that are not shared is regarded as an indication of dissimilarity" (Watanabe 1969, pp. 377-8).
- "[a]ny two objects, in so far as they are distinguishable, are equally similar. A corollary to this theorem is that if we want to revive the idea of class of similar objects we have to ponderate (give weights to) the predicates so that we can say that in order for two objects to be similar to each other they have to share more important (weighty) predicates" (Watanabe 1986, p. 7).

In a moment, we'll look at a elaborated illustration of Watanabe's approach, but before turning to this it is important that we recognize two things: First, the fundamental purpose of his theorem is to disclose and to underscore that purely formal methods alone can never provide a basis for the judgments that we make of similarity and dissimilarity. He expresses this by saying that such judgments must rely on "extra-logical" weighting, which therefore puts similarity judgments beyond the *possible* justification they might receive from the abstract, universally applicable principles we associate with logic and mathematics. In this sense, the Theorem of the Ugly Duckling comprises yet another variety of limitation that formal systems cannot escape (others include, for example, Gödel's proofs and the Löwenheim-

---

[11] To be accurate, actually a collection of several theorems that are linked together.

[12] The original text reads: "... *qu'il est impossible de parler de la similarité et de la dissimilarité des objets tant que l'on essaye de définir ces notions sur la base des nombres de tous les prédicats possibles auxquels les objets satisfont ou ne satisfont pas*" (Watanabe 1965, p. 49).

Skolem Theorem, which we will meet in the next section).

Secondly, we need to recognize that the Theorem of the Ugly Duckling does not undermine judgments of similarity or commonality, but instead it brings to light the essential framework-relativity of all such judgments. This second aspect of Watanabe's result was perhaps not recognized by him and was not developed in his publications; we'll discuss this aspect later.

Here, I would like to help the reader see the strength and generality of Watanabe's theorem by means of a simple illustration that has intuitive force, from which it is possible to infer Watanabe's result in a semi-formal way.

We have seen that judgments of commonality presuppose recognition of salient properties; human attention tends to be selective and—at this stage of our knowledge, for lack of a more satisfying explanatory term—human attention appears to be innately discriminatory, placing objects of perception and their properties in intuitive or instinctive classes. People frequently make such comparative judgments two objects at a time. If we wish to analyze the logic underlying judgments of commonality, it is helpful to do this on a greatly simplified basis by focusing on comparative judgments that do exactly this, that relate two objects at a time. (We recall that this approach was suggested many years ago by Gulliksen (1947).) As did Watanabe, we'll follow that pair-by-pair approach here.

Let us consider a moderately complex situation that illustrates the possible combinations that four objects may have when there is a designated property they may possess. (This was not, incidentally, Watanabe's more generalized approach, but it may serve to illustrate what he demonstrates.) We wish to see how often any pair of the objects has, or fails to have, that property in common; when two objects both fail to have a individual property, they also have that fact in common—and hence have in common the property that expresses that fact: They then share the property of not having that particular property. This is not far-fetched: Consider the property of having citizenship. Not having that property is to be stateless, which is certainly a property that people can have and some do.[13]

We wish to tally the total number of possible property commonalities the four objects may have. In the table that follows, the four objects are named A, B, C, and D. The four vertical columns on the left below the letters A, B, C, and D enumerate all of the possible 16 numbered combinations of properties that the objects may have (the combinations are exhaustive; there are no others): If an object has the property, this is designated by a '1'; if it does not have that property, a '0' shows this. Line 1 of the table is the case where all four objects possesses in common the property designated by '1'; the last line of the table represents the case where the four objects all fail to have that property.

Suppose now that we ask how often, for the 16 possible combinations of properties, any pair of the four objects has properties in common, or fails to have

---

[13] Citizenship and statelessness may appear at first sight to be unusual as complementary properties, but only because we are not accustomed to thinking of non-possession of a given property as itself a legitimate property.

Here is Watanabe's way of expressing this: "... *la négation simultanée d'un prédicat par deux objets est une indication de similarité aussi significative que l'affirmation simultanée*" (Watanabe 1965, p. 51) [... the simultaneous negation of a predicate by two objects is as much an indication of their similarity as is the simultaneous affirmation (of that predicate)].

those properties in common—i.e., we want to identify all combinations for which a given pair of objects possesses 1's or 0's in common.

As the table shows, there are 6 possible object pairs: A & B, B & C, C & D, A & C, A & D, and B & D. We place a check mark in the table whenever a specified pair of objects has 1's or 0's in common. We do this to make evident how often, out of the 16 possible combinations of properties, commonalities show up. For example, on line 7, we note that column B and column C both have the same value; both are 0s; and so we place a check mark below column B&C on line 7. In the same way, we see that both columns A and D have the same value; both are 1s; and so we also place a check mark below column A&D on line 7. By doing this, Table 1 visually represents all the ways in which four objects, when analyzed pair-by-pair, may have properties in common.

| | A | B | C | D | A&B | B&C | C&D | A&C | A&D | B&D |
|---|---|---|---|---|---|---|---|---|---|---|
| 1 | 1 | 1 | 1 | 1 | ✓ | ✓ | ✓ | ✓ | ✓ | ✓ |
| 2 | 1 | 1 | 1 | 0 | ✓ | ✓ | | ✓ | | |
| 3 | 1 | 1 | 0 | 1 | ✓ | | | | ✓ | ✓ |
| 4 | 1 | 1 | 0 | 0 | ✓ | | ✓ | | | |
| 5 | 1 | 0 | 1 | 1 | | | ✓ | ✓ | ✓ | |
| 6 | 1 | 0 | 1 | 0 | | | | ✓ | | ✓ |
| 7 | 1 | 0 | 0 | 1 | | ✓ | | | ✓ | |
| 8 | 1 | 0 | 0 | 0 | | ✓ | ✓ | | | ✓ |
| 9 | 0 | 1 | 1 | 1 | | ✓ | ✓ | | | ✓ |
| 10 | 0 | 1 | 1 | 0 | | ✓ | | | ✓ | |
| 11 | 0 | 1 | 0 | 1 | | | | ✓ | | ✓ |
| 12 | 0 | 1 | 0 | 0 | | | ✓ | ✓ | ✓ | |
| 13 | 0 | 0 | 1 | 1 | ✓ | | ✓ | | | |
| 14 | 0 | 0 | 1 | 0 | ✓ | | | | ✓ | ✓ |
| 15 | 0 | 0 | 0 | 1 | ✓ | ✓ | | ✓ | | |
| 16 | 0 | 0 | 0 | 0 | ✓ | ✓ | ✓ | ✓ | ✓ | ✓ |
| **Total checked** | | | | | 8 | 8 | 8 | 8 | 8 | 8 |

**TABLE 1**

If we add up the check marks in the column A&B, we see there are 8 total. And we see there is a total of 8 check marks for *every possible pair of objects*. But it is equally important to recognize that there are an equal number of times when pairs of objects *fail* to have properties in common; there are also 8 *unchecked* cells in *every* column below the possible pairs of objects.

The reader should be able to generalize inductively from this illustration that involves 4 objects: As he or she can verify, when there are 5 objects, the number of property permutations increases from 16 to 32, the number of checked commonalities increases from 8 to 16, while the number of unchecked cells continues to remain equal to those that are checked. No matter the number of objects, there will be an equal number of times when pairs of those objects have properties in common as there are times when they do not.

The above simplified example is hopefully not too simple that the reader will dismiss what it conveys. Watanabe provides proofs for the general case in which $n$ may be any number of objects, and therefore $2^n$ represents the possible combinations of property pairs they may have.[14]

The Theorem of the Ugly Duckling tells us something about the occurrence of commonalities that, from a strictly logical point of view, is surprising to many people: If you take any two objects at random, the number of property commonalities they can possibly have is equal to the number of commonalities that two objects that are judged to be similar may possess. Hence, an arbitrarily chosen object—a duckling (or a stone, for that matter)—when compared with a swan have just as many properties in common as do two swans. Alternatively expressed, there are just as many ways for two objects that are judged to have properties in common to fail to have other possible commonalities as there are ways in which two objects judged to be dissimilar also have no properties in common. The number of properties that two swans (which are believed to have properties in common) possess in common is no more, no less, than the number of ways they fail to have properties in common, and the latter number of failed commonalities is the same as in the case of two objects considered to be unrelated.

Now, what is the point of this formal analysis? Clearly, in ordinary experience we can and do recognize when two objects have something in common, and we can and do recognize when two objects don't. The purpose of Watanabe's Theorem of the Ugly Duckling is to make evident that, no matter how many properties may be distinguishable, and no matter how many objects we may wish to compare that have or fail to have those properties, any commonalities that we end up paying attention to are *not* commonalities that purely formal, logical criteria determine. Instead, logical criteria alone leave us in a state of *necessary* ambiguity whenever judgments of commonality are desired. To progress beyond this *inescapable condition of ambiguity*, we are forced to specify and then identify in objects properties that, *for extra-logical reasons*, are considered to be important. It is only relative to those properties that are elevated in *stipulated* importance that objects are ever justifiably considered to have anything in common.

Finally, in order to stipulate which properties are to be considered important, we must apply some criteria or standards of importance, and it is this *choice* of criteria or standards, based on considerations that no formal considerations in mathematics can settle, that we so often mistakenly believe point to the "real, actually existing, natural kinds" which traditional taxonomists thought they were identifying when they classified distinct species.

Earlier I referred to a paper by Nelson Goodman that is directly relevant to a clearer understanding of the species problem. Before we leave the present discussion of the Theorem of the Ugly Duckling, I'd like to summarize several observations made by Goodman that relate to Watanabe's result, although Goodman was apparently unaware of it. Goodman's paper on similarity has come to the attention of a few biologists (a fairly long discussion of Goodman appears, for example, in Medin, Goldstone, & Gentner (1993)—and yet the authors only provide in passing—even so, a great rarity in taxonomic literature—a single and unexplained citation to Watanabe's more substantive work).

---

[14] Watanabe provides proofs, for example, in Watanabe (1965; 1969, §7.6; 1985, Chap. 4).

In his paper, "Seven Strictures on Similarity," published more than a decade after Watanabe's first presentation of his Theorem of the Ugly Duckling, Goodman refers to Gulliksen (1946), whose method of paired comparisons we met earlier, but, as we know, not to Watanabe, whose result Goodman comes very close to duplicating, although he does this by means of a less mathematically rigorous route. In his paper, Goodman sets out to critique the concept of similarity and to point out certain of its limitations. To do this, he formulates a group of "strictures," or injunctions, against the widespread uncritical use of the concept of similarity; he argues for these strictures informally, and unlike Watanabe does not provide any proofs.

The last of these strictures parallels Watanabe's result; as Goodman expresses it: "Similarity cannot be equated with, or measured in terms of, possession of common characteristics" (Goodman 1972, p. 443). Here is the core of Goodman's argument:

> When, in general, are two things similar? The first response is likely to be "When they have at least one property in common." But since every two things have some property in common, this will make similarity a universal and hence useless relation. That a given two things are similar will hardly be notable news if there are no two things that are not similar.
>
> Are two things similar, then, only if they have all their properties in common? This will not work either, for of course no two things have all their properties in common. Similarity so interpreted will be an empty and hence useless relation. That a given two things are similar in this sense would be notable news indeed, but false.
>
> By now we may be ready to settle for a comparative rather than a categorical formula. Shall we say that two things $a$ and $b$ are more alike than two others $c$ and $d$ if $a$ and $b$ have more properties in common than do $c$ and $d$? If that has a more scientific sound and seems safer, it is unfortunately no better, for *any two things have exactly as many properties in common as any other two* [italics added]. If there are just three things in the universe, then any two of them belong together in exactly two classes and have exactly two properties in common: the property of belonging to the class consisting of the two things, and the property of belonging to the class consisting of all three things. If the universe is larger, the number of shared properties will be larger but will still be the same for every two elements. Where the number of things in the universe is $n$, each two things have in common exactly $2^{n-2}$ properties out of the total of $2^n - 1$ properties; each thing has $2^{n-2}$ properties that the other does not, and there are $2^{n-2} - 1$ properties that neither has.[15] If the universe is infinite, all these figures become infinite and equal....

> More to the point would be counting not all shared properties but rather only important properties—or better, considering not the count but the overall importance of the shared properties. Then *a* and *b* are more alike than *c* and *d* if the cumulative importance of the properties shared by *a* and *b* is greater than that of the properties shared by *c* and *d*. But importance is a highly volatile matter, varying with every shift of context and interest, and quite incapable of supporting the fixed distinctions that philosophers so often seek to rest upon it. (Goodman 1972, pp. 443-444)

The history of the species problem, with its many competing, alternative definitions of species, has tended to confirm Goodman's generalized conclusion—that judgments of similarity are simply too "volatile" and incapable of supporting even taxonomists' distinctions. Efforts to find a solution to the species problem seem to be petering out—unfortunately not as a result of an awareness of compelling, theoretically provable reasons—but due to the impatience, frustration, and resulting emerging awareness among taxonomists that controversies over the definition of species can ever be settled.

Criteria of importance, which enable taxonomists to divide up the biological landscape into classifications of similar and distinct types of organisms, are, to adopt Goodman's way of expressing this, themselves highly volatile and vary with context and interest—but are they, in the final analysis, incapable, as Goodman thinks, of supporting the fixed distinctions that taxonomists so often seek to rest upon them? If there is no justifiable unique and compelling system of classification? are individual taxonomies therefore arbitrary because—from a rigorous, strictly logical point of view—each fades, so to speak, into an inescapable fog of ambiguity?

The conclusion reached by Watanabe is simply stated: Comparative judgments of similarity, and the classifications they may lead to, are fundamentally based on our prior acceptance of criteria of importance, i.e., those criteria that specify what it is we are to attend to. This prior acceptance cannot, as we have seen, be given a formal, logical, mathematical justification; it is an expression *only* of needs, interests, and utilities that must come from an extra-logical origin. As Watanabe noted in connection with the Theorem of the Ugly Ducking: "A corollary to this theorem is that if we want to revive the idea of class of similar objects we have to ponderate (give weights to) the predicates so that we can say that in order for two objects to be similar to each other they have to share more important (weighty) predicates" (Watanabe 1986, p. 7). Once criteria of importance have been specified, recognition of commonalities *is* made possible, and this will happen in a manner that avoids the form of ambiguity which strict formal methods of and by themselves cannot escape.

---

[15] Goodman's calculations here may be somewhat hard to follow. He would apparently like the reader to assume that the "things" he is referring to are "bare logical particulars"—that is, objects of reference that possess no individuating properties in the concrete sense: These "things" are merely, and only, numerically distinct. Their "properties" therefore consist of nothing more than properties of membership in distinguishable classes to which they may belong.

(At this point, we may be reminded of the theoretically impossible task set by the Rand-NSF project described earlier, which wished to develop general algorithms capable of detecting commonalities—but without first inputting specified criteria, stipulated ahead of time, by means of which to identify those commonalities. We now recognize just why that prior input is indispensable.)

We are left unavoidably with what I have elsewhere termed "framework-relativity":[16] Just as the routinely employed coordinate system establishes a framework relative to which physical measurements have significance, so does the specification of criteria of importance provide the basis for a pattern recognition framework relative to which the identification of commonalities can take place.[17] In a later section, we'll discuss framework-relativity in somewhat greater detail. Here, to complete the account that can be given in this monograph of the logic of commonality as it relates to the species problem, we turn to a second formal result.

## 4.4 *The Löwenheim-Skolem Theorem*

There are two main conceptually problematic aspects of the controversy surrounding the definition of species: The first, which we have now examined, relates to the logic underlying the recognition of commonalities among objects; the second concerns the relationship between taxonomies and the objects they classify. The logic of commonality recognition throws a certain amount of light on the relationship between theories of classification and the universe of natural organisms. The light that it sheds has shown that the naturally occurring range of organisms does not of course come already divided up into pre-labeled natural classes, but rather, as we've seen, such classes require for their potential recognition a prior specification of salient features determined by extra-logical criteria of importance.

The relationship between theory and its intended referents can be conceptualized in a highly abstract way, greatly removed from any reference to specific theories and their corresponding applications. This makes it possible to step back and appraise our thinking so as to clarify what can, and what can not, be expected from the theories we formulate in our efforts to represent the structure of a given range of objects and their interrelationships.

---

[16] See footnote 1.

[17] Goodman would disagree: Where, for him, a physical reference frame does successfully eliminate ambiguities of measurement, specifying what I have called criteria of importance cannot, in his view, be similarly successful. If I've understood him correctly, Goodman reasons that once we have identified a set of criteria of importance relative to which similarities are recognized, we have either assumed in a circular fashion what we wished to prove, or there is nothing about similarity that we can explain other than simply to re-state that two things are similar because they are similar in such-and-such respects: "similarity tends under analysis either to vanish entirely or to require for its explanation just what it purports to explain" (Goodman 1972, p. 446). This reasoning is loose and for me unconvincing. Both a physical frame of reference and, e.g., a perceptual frame of reference achieve their ends: The physical frame of reference that establishes an origin relative to which measurements can be made is comparable to the perceptual frame of reference that specifies a matching criterion relative to which commonalities can be identified. Both frames of reference remove the relevant ambiguities.

The area of mathematics that deals with this subject is model theory. Within model theory, the proof most relevant to this monograph is a result that has come to be known as the Löwenheim-Skolem Theorem. This mathematically technical achievement is, as far as I am aware, never mentioned in discussions of the species problem. The closest that anyone has come to relating this major theorem to the general area of our discussion was philosopher Hilary Putnam (1981), as we shall see in a moment.

The Löwenheim-Skolem Theorem represents the accumulated result of work by mathematicians Leopold Löwenheim, Thoralf Albert Skolem, and Anatoly Ivanovich Maltsev over a period of years from 1920 to 1936. Its principal significance is its contribution to model theory, which, as an independent area of study in mathematics, is not directly relevant to our subject-matter here. However the theorem does yield informative consequences for a highly general understanding of the relationship between theories—here, we have taxonomies specifically in view—and the ensembles of objects those theories are about—here, the wide range of naturally occurring organisms.

A few basic concepts are needed in order to state some of the consequences of the Löwenheim-Skolem Theorem as they apply to taxonomic theory. In mathematics, model theory is encountered in a study of formalized systems. When it is possible to establish a correspondence, called a mapping, between the elements of a given formalized system (e.g., sets, functions, relations) and a domain of objects so that the theorems of the formalized system hold true for those objects, their properties, and their relationships, then that domain of objects is said to constitute a *model* of the given system. Expressed differently, a model comprises an *interpretation* of the formalized system such that each of the system's theorems holds true in that model.

It is possible to understand the Löwenheim-Skolem Theorem in two distinct ways: There is the level of technical understanding that is reached in the mathematical study of model theory, and there is an extended level of understanding that applies the formal result beyond its strict mathematical development. When we resort to an extended understanding of this kind, on the one hand we run the obvious risks of misapplying and over-generalizing a mathematical result. On the other hand, legitimate extended applications of a formal result can provide insight we might otherwise not gain.

Here, this not being a study of model theory, a restatement by extension of the Löwenheim-Skolem Theorem will be sufficient:[18] Informally expressed, the theorem proves that, for any formalized system beyond a certain very minimal degree of complexity,[19] there will always be a multiplicity of alternative models of that system. In other words, for any such basic formal system, that system's own formal constraints—the formal axioms and rules that define the system—do not force a single, unique interpretation. As a consequence, such formal systems lead—in a way unlike and yet parallel to the logic of commonality that we have discussed—to an *inescapable ambiguity of interpretation*: It is logically not possible to constrain such a formal system to one designated, unique model. There will always be alternative, unintended interpretations of any such system.

---

[18] Stated more strictly, the theorem is: If a formal system $S$ is a first-order theory, then the theorem shows that when $S$ has a model, then $S$ has a model with a denumerable domain. A strong form of the theorem proves that if $S$ has an infinite model, then it also has models of every transfinite cardinality. Since no two models of different cardinalities can be isomorphic, $S$ has incompatible models.

Putnam has been one of comparatively few theorists who have recognized a more widespread application of the Löwenheim-Skolem Theorem as understood in this extended sense. Here is the way he expressed this: "[I]t is possible to interpret ... [the sentences of an] entire language in violently different ways, each of them compatible with the requirement that the truth-value of each sentence in each possible world be the one specified" (Putnam 1981, p. 33). —In other words, once a formal system possesses a certain minimal degree of complexity, there will be an endless number of models that "fit" the system so that all of the system's theorems hold true in each of the models, and yet many of these models will be incompatible with one another. In this sense, it is always possible to interpret such a formal system in "violently different ways."

There is an important "reverse" way in which to understand this result. We may begin with a domain of objects that is minimally complex so as to require for its representation a formal system of the sort we have been discussing. We may then ask whether a reverse application of the Löwenheim-Skolem Theorem holds in such a case—whether, in other words, there will be indefinitely numerous ways in which to represent that domain of objects through the expressive means of distinct and often incompatible formal systems.

This is indeed the case, for once the domain of objects has been formally expressed by a system *S* so that the theorems of *S* are true of that domain of objects, then the Löwenheim-Skolem Theorem immediately associates with *S* the inescapable ambiguity of an endless number of alternative systems to *S*, many of which will not be compatible with *S* or with one another. In short, if we begin with a model of sufficient complexity—let us say, the domain of naturally occurring organisms—then *there not only is no single unique theoretical representation of those objects, but there cannot be*.

## 4.5 *Taxonomic Consequences of the Logic of Commonality and of the Löwenheim-Skolem Theorem*

> *The fact is that species are hard to identify for a variety of reasons related to the various ways that they can be truly indistinct, and no criterion that presumes to delineate natural boundaries can overcome this. It seems possible that this lesson is beginning to sink in. Perhaps our fifty-plus years of*

---

[19] Predicate logic supplemented so as to include arithmetic.

*argument over multiple species concepts has been a necessary prerequisite for this realization.*

— Hey (2006, p. 449)

Jody Hey, in her paper quoted above, attributes "the failure of modern species concepts" (the title of her paper) to the inherent fuzziness of species boundaries, the inadequacy of any species identification criterion that can overcome this, and the passage of more than fifty years of inconclusive debate over alternative species definitions. These are no doubt important factors to be mentioned; however, the very existence of the species problem, as I hope this monograph gradually makes evident, is not due primarily to fuzzy boundaries, lack of adequate recognition criteria, or human exhaustion after long and inconclusive controversy. The source of the problem is rather to be found in a lack of explicit understanding of the logic underlying the identification of commonalities, and in a lack of awareness of the intrinsic limitations of theories in relation to unstructured information.

At this point, we need to review the results we have reached in light of their direct application to the species problem. We are reminded that a given taxonomy is a system of classification that sorts the objects it classifies into differentiable classes. And as we have seen, a dominant concern in the history of taxonomy has (with the exception of the subjectivists, considered later) been to produce a system of classification that is both *real* (not the artifice of human subjectivity), and *univocal*—in short, a taxonomy that "cuts the natural world at its naturally occurring joints," that identifies naturally occurring classes of organisms in a non-arbitrary way that is unique so that general assent to that system of classification is compelling. The classical example of the periodic table of the elements is often cited as a desirable paradigm of these goals; its system of classification of the elements is perceived as so systematic, precise, and informative of reality that the system has gained universal acceptance.

I have discussed in some detail two main formally limitative results, Watanabe's Theorem of the Ugly Duckling and the Löwenheim-Skolem Theorem. The first shows that it is logically impossible to capture the meaning of commonality (or similarity) by strictly formal means. From a logical point of view, the recognition of commonalities brings with it logically inescapable ambiguity. The second theorem shows that the relationship between any minimally complex formal theory and the domain of objects to which it can be applied truly is never a unique relationship, but also brings with it logically inescapable ambiguity. As a result, we are forced to the realization that the recognition of commonalities entails reliance upon extra-logical criteria that must be stipulated, while the formulation of any formal theory, such as a formalized system of classification in taxonomy, can never—in principle—be univocal (as for having "real" reference as opposed to being a "subjective construction," we will come to this later).

The applicability of these two mathematical theorems to the species problem can only take us so far, as the reflective reader will no doubt have perceived. For one thing, biologists do not tend to formalize their taxonomies, and so they might wonder whether formal limitative results apply to their work. For another, few biologists are concerned with the question whether their identification of species can be theoretically formulated in such a way as to lend itself to computerized pattern recognition; that is, they are not often concerned to formulate formal algorithms capable of differentiating and identifying species.

Both of these objections are, when taken at face value, completely justifiable. Right now, taxonomies are seldom expressed in the form of formalized theories, and there is at this time no widely felt need in connection with species identification to enable machines to simulate human pattern recognition abilities (although this is changing with the rapid development of computer- and even chip-enabled microorganism species identification). Let us, nonetheless, suppose that taxonomists will not experience a need to formalize their systems of classification, and will have no need to develop algorithms to recognize and discriminate among species. In this case, what may be learned from the formal results?

In a sense, this is a somewhat like asking whether Catalan architect Antoni Gaudí should have learned the mathematics he did not know when designing the famous Templo Expiatorio de la Sagrada Familia in Barcelona—after all, he was able get along without the underlying math by visualizing and measuring the distribution of forces in the complex structure of the cathedral by means, instead, of an analog system of hanging lead weights from strings.

Knowing the underlying mathematics does, of course, provide a step-up in one's level of understanding. But, more than this, the Theorem of the Ugly Duckling and the Löwenheim-Skolem Theorem both make us aware of the central role of what I have called "extra-logical" factors that, in principle, cannot be derived from strictly logical grounds.

Had it been known that these factors were indeed *necessarily* extra-logical in origin, and of course had this knowledge been accepted and acted upon by biologists, the species problem—as a search for the one, universal, compelling definition of species— would simply not have happened.

Are we then left to pass judgment that the many alternative species definitions are arbitrary, mere artifacts of human subjectivity, that the species distinguishable by their means are "invented" rather than "discovered"? Are the definitions that have been offered therefore not "objective" and so fail to "correspond to reality"? As we shall see, the answer to each of these questions is no.

## 5. The Illogic of Ontological Bias

The species problem constitutes an unsolvable problem if by a species one means a classification that is real, natural, and independent of human theoretical construction. Such a conception of species embodies within it a strong ontological bias of realism, and this bias fundamentally expresses a demand that cannot, in principle, be met.

In previous sections, we have encountered ontological bias in two similar contexts: Early in this monograph we saw that when multiple observers agree in their recognition of a pattern, and a consensus of judgment is established, then that pattern tends to be considered "real" and "objective," while when an individual who sees a pattern where other people do not, as in apophenia and pareidolia, then that pattern is judged to be a "subjective expression of imagination." In a later section we saw that when people are presented with unstructured information in relation to which a pattern becomes recognizable, the general human tendency is both to believe the pattern "was there all along," and to judge that pattern to be

"real," and therefore "discovered" and not "invented."

The ontological biases that affirm realism and objectivity, on the one hand, and subjectivity and imagination on the other, are seldom subjected to critical analysis, and more rarely still when a set of objects of reference is identifiable only from the standpoint of a particular theoretical framework. To begin such an analysis, let us distinguish these ontological biases according to whether they are *perceptually-based* or *theoretically-based*.

The logic underlying the pattern recognition of commonalities is especially instructive in bringing out the *illogic* of many ontological biases. By now, it should be clear that there is no such thing as pure pattern recognition—i.e., pattern recognition that proceeds in a vacuum, so to speak, that does not rely upon previously specified criteria that make it possible to identify those patterns. Pattern recognition in this sense is always framework-relative, relative to a framework in terms of which required criteria of identification are specified. As we have seen in explaining Watanabe's theorem, pattern recognition does not occur, and indeed *cannot* occur, independently of a framework that specifies what properties or features are to be selectively sought and identified.

The variety of pattern recognition that has been the primary focus of this monograph is perceptually-based since the patterns recognized by species definitions are those perceived by human biologists. The ontological biases that have found their way into discussions of pattern recognition also originate in a perceptual context. We have described the logic that must be granted in order for it to be possible to detect commonalities shared among objects. And we have seen that this underlying logic rules out the possibility that such commonalities can be recognized independently of a prior specification of criteria of importance.

Now, in this context, a subjectivist's ontological bias claims that a recognized pattern originates in subjective imagination, while a realist's objective bias asserts that the pattern has an independent reality of its own. The subjective bias claims that the pattern is "generated by imagination" (alternatively, is "formulated as a theoretical construct"); the objective bias claims the pattern "was there all the time before it was recognized." Both of these claims, as we shall soon see, trespass beyond the boundaries of possible reference: Once we recognize that pattern recognition is framework-relative, this should be the end of the matter. But it so often is not; instead, ontological claims are made for which we'll find that justification is, in principle, impossible.

To move to a more inclusive level of generality, when a specific theoretical framework comprises the basis for reference to a given group of objects, the identification of those objects—and indeed their very identity as understood by means of that framework—is made possible by that frame of reference. It does not, and again it cannot, make sense to pretend as though such objects have an autonomous existence apart from the framework that *permits* reference to them.

Consider a simple example of a Cartesian coordinate system with three axes, *x*, *y*, and *z*. We use the system to identify a specific point, e.g., the point whose coordinates are $(1, 5, \sqrt{2})$. The "independent existence" of that point is a question that does not arise; it would make no sense. The point we have identified relies for its identity upon the coordinate system we presuppose when making reference to that point. The two are functionally tied together; with a formal coordinate system, we tend to see this immediately, and it presents no problem. But when we transfer our attention to other more complex, richer frameworks of reference—for example, the framework of sense perception, or of conscious awareness, or of language—matters also become more complex, and they become correspondingly muddled.

Earlier, I discussed the Löwenheim-Skolem Theorem in its application to those models for which the theorems of a formal system are true. This meaning of the term 'model' is its more technical sense: a model being the interpretation of a formal system for which its theorems hold true. There is also a less technical sense of the term, meaning the range of objects to which we wish to refer. The general nature of ontological bias can be made clearer by employing this second sense.

Ontological bias can then be said to relate to the question whether the models to which our theories apply can be referred to as autonomous, that is, without relying upon those theories. When we question ontological bias, at a most basic level we are asking whether the domain of objects to which we wish to refer *can be referred to* as having an *autonomous* existence—i.e., separable and independent, *known without employing the means necessary to refer to those objects*. The age-old quandary, whether a falling tree in the forest makes a sound if no one is there, is of this kind: It involves a camouflaged attempt at one and the same time to refer to a set of objects (a tree that falls among others in a forest), while taking away the framework in terms of which trees, forests, and the noise of falling trees can possibly be recognized.

Of the same kind, as an example offered here for philosophers, is the Kantian claim that the mind actively structures incoming perceptual data so that the data take spatial and temporal form for us. Again, there is a camouflaged attempt at one and the same time to refer to spatial, temporal objects and yet to hypothesize an impossible frame of reference that would permit us to recognize that these perceptual objects (which in the Kantian view must necessarily be spatial and temporal in form) are the mind's genetic production constituted from "not yet processed incoming raw data." If to become conscious of what we think of as physical objects *is* to be conscious of them as spatial and temporal, then we cannot possibly become conscious of a hypothesized genetic activity that begins with "raw data," "processes the data," and results in the objects of which we are conscious. The system of reference that would be necessary to do this is precluded by the very framework in which it is claimed that the mind does this automatic constitutive processing.

Realists who claim that objects exist independently of the reference frames in terms of which those objects are identified, and subjectivists who claim that objects derive their form, structure, and nature from human subjective activity, are both victims of *projective thinking*. Realists, because they seek to refer to an ontological autonomy that cannot be reached through any referential means, and subjectivists, because they, too, attempt to refer to a genetic, productive basis whose functioning cannot, in principle, by their own contentions, be evidenced.

The realist, in claiming that a domain of objects is "discovered," seeks to refer in a manner that exceeds the limitative capacity of his or her system of reference—no matter what that system is. He or she cannot, in principle, refer to that which lies beyond the possible boundaries of any frame of reference, and by doing this justify belief that a domain of objects "exists independently." The subjectivist, in claiming that a domain of objects is "invented" rather than "discovered," similarly has no possible way to justify his or her claim; like the realist, he or she seeks to refer in a manner that exceeds the limits of anyone's referential capacity; there is no possible way to justify the alleged fact that, for example, real numbers would not exist were it not for the alleged constructive activity of the mathematician.

There is a deeply rooted and virtually universal propensity among people to attempt to *project* beyond the boundaries of their frames of reference in an effort to endow the objects of their projected beliefs with autonomous, or, alternatively, subjectivity-dependent, existence. We see this in virtually all areas of human experience and endeavor. Even in mathematics, the most formally self-conscious discipline, Platonist mathematicians still believe that such things as the real numbers have an autonomous existence, a reality independent of human minds and the systems of mathematical reference that are relied upon by those minds in order to refer to the domain of real numbers. In a parallel fashion, subjectively-oriented intuitionist mathematicians continue to believe that the domains of objects to which mathematics refers owe their genesis and their existence to the constructive activity of human minds—yet to maintain this in any meaningful way requires a referential standpoint that is denied to any human mind that would assert those objects' dependency upon that mind; the conditions for referring to those objects as mind-dependent are conditions that would need to be suspended (in order to know that they are mind-dependent), which they cannot.

Neither realists nor subjectivists have—or *can* have—grounds on which to stand. Elsewhere, I have given detailed proof of this proposition,[20] but for purposes here it will be sufficient if the reader is willing to weigh the persuasive force of adopting general framework-relativity against its rejection:

To accept general framework-relativity is to realize and to respect the intrinsic connectedness to an implicit background or explicit frame of reference of all forms of human recognition of patterns, of reference to objects of whatever kind, whether concrete or abstract, of identification of any possible object of attention, no matter what level of discourse or level of generality of theory.

In contrast, to reject framework-relativity is to countenance beliefs and assertions that claim the framework independence—the autonomous reality or independent meaning—of what is believed or asserted. This claim to autonomy is made in two steps: by insistently trespassing beyond the limits of the reference frames that are relied upon in order to articulate those beliefs and assertions, and at the same time by employing those very reference frames in the very act of denying that one does this.

We have a variety of metaphors that capture in a picturesque way the illogic that is involved: A century and more ago, the apt phrase would have been "hoisted by your own petards"— to be blown up by your own bomb;[21] today, we're more inclined to speak of "pulling the carpet out from under one's own feet." What blows up in one's face, or pulls one's feet out from under one, is not the result of a factual dependency upon a *particular* framework. It is rather what I have elsewhere called a *metalogical* relationship:[22] Framework-relativity means essential reliance on *some* appropriate frame of reference that makes desired forms of reference possible, and it is precisely *that* reliance which is denied by projective assertions and beliefs.

---

[20] See footnote 1.

[21] See, e.g., the author's "Hoisted by their own petards: Philosophical positions that self-destruct," Bartlett (1988).

# 6. The Solution of the Species Problem

In §2, I distinguished three varieties of definition relevant to an understanding of the species problem: stipulative, real, and coordinative definitions. We have seen how fundamentally important stipulative definitions are in providing the basis for the pattern recognition of commonalities: They specify criteria of judgment concerning the features or properties that are to be selected for the recognition of discernible classes of organisms. Beyond this, on the one hand, realist species definitions seek to function as real definitions that have empirical, mind-independent, informative content concerning the classes of organisms that are to be distinguished from one another in a taxonomy; on the other hand, subjectivist species definitions deny they have a basis in reality and therefore assert they are only human constructions. Together, and despite sometimes differing ontological biases, the stipulative, realist or subjectivist, and coordinative functions of any species definition work together in systematically organizing salient features of organisms in a wholly framework-relative manner.

At the beginning of this monograph, the radical claim was made that, once we recognize the constraints that apply to the logic underlying species definition, it is possible to define species successfully in a way that is conceptually iron-clad, that is, cannot *not* be accepted without inconsistency.

How, given our acceptance of thoroughgoing framework-relativity, is this possible?

It is possible precisely thanks to the interaction between the logical constraints of commonality recognition and framework-relativity. We have seen how the detection of commonalities is dependent, in principle, upon the logically prior specification of properties judged to be important. Once a taxonomic theory is formulated and the properties that are to be recognized are specified, the application of the theory to the intended range of naturally occurring organisms results in a system of classification, a taxonomy of species. We recognize that this framework-relative identification and classification of organisms can both be informative, and in this way embody the empirical content intended by a so-called real definition, and yet the distinguishable species which the taxonomic theory defines derive their sense and applicability wholly in a theoretically-relative manner.

Here lies, in these few summary statements, the solution of the species problem. We have been led to this solution, in a series of theoretically compelling steps, by an examination of the logic underlying the recognition of patterns of commonality and by an understanding of the framework-relativity of reference.

Several conclusions follow: Species, so-defined, are neither independently existing objects of reference, nor are they subjective constructs. To claim either is to engage in projective thinking, and thereby to pull the carpet out from under one's own feet. *Species, understood as framework-relative, are neither "discovered" nor "invented"—again, their definitions derive their meaning and their applicability only in terms of the coordinative frame provided by a given taxonomy.* A definition of 'species' establishes a frame of reference, a context in which a domain of objects can be systematically classified. Species that are so defined derive their identity, their very recognizability, with respect to that referential context. *There is no subjective or objective, realist or subjectivist, ontological bias involved.* We saw this immediately in connection with the essential relativity of a specified coordinate to its appropriate coordinate system; it is no different here.

---

[22] Bartlett (1971, 1975a, 1976, 1980, 1982, 1983, 1987, 1992).

Understood in this conceptually self-accountable way, a definition of species cannot *not* be accepted *given* the coordinative framework which that definition presupposes.[23] This is a tautologous assertion, but it is not circular reasoning that is logically vicious. The assertion follows from an understanding of the logic underlying all pattern recognition, and from due respect for framework-relativity. Once we agree to respect both the underlying logic and framework-relativity, the tautology follows.

A species, then, is exactly what a given definition claims it is. There *can be* no unique, univocal definition. To wish for this, is to wish for the theoretically impossible; to be disappointed that no "better, all-embracing" definition can be formulated, is to misunderstand the task. Alternative definitions of species form an inescapable plurality, and, in different ways, pick out sets of salient features in terms of which one species, within that system of classification, is to be distinguished from others. In order to escape from the unavoidable ambiguity that accompanies any attempt to recognize commonalities on a purely logical basis, and as the Löwenheim-Skolem Theorem requires, we are forced to accept a plurality of taxonomies, each of which, in and on its own terms, cannot *not* be accepted once it is formulated in terms of specified criteria that are necessary in order to detect the commonalities which that system of classification recognizes as a basis for species identification.

This conclusion leaves us unavoidably with what, for many biologists, is an inherently unsatisfactory solution of the species problem. For them, a solution is unsatisfactory if it does not lead to the result they desire, namely, a definition of species that is univocal, universal, theoretically compelling, and, if they are realists, revelatory of "real, independently existing classes." The Theorem of the Ugly Duckling shows that the recognition of commonalities is logically arbitrary, subject to inescapable ambiguity, and that such commonalities cannot, in principle, be detected without prior reliance upon extra-logical criteria of judgment. Framework-relativity shows ontological biases to be not merely biases, but expressions of self-undermining logical error. And the Löwenheim-Skolem Theorem demonstrates that, for any minimally complex domain of objects, there will be indefinitely numerous alternative theories which can truly account for that domain, and that many will do this in incompatible ways.

---

[23] There are two senses in which this is the case: (i) from the standpoint of the underlying logic this monograph has sought to describe, and (ii) from the standpoint of an individually formulated taxonomic system. In the first sense, the theory of species definition developed in this monograph cannot not be accepted without logical inconsistency; in the second sense, once the identification criteria proposed by a particular definition of species are granted, the species classifications which that taxonomy recognizes cannot not be accepted without begging the question or incurring the charge of pointless objection (see §7.3).

For taxonomists who have been committed to the search for a univocal system of classification—whether for a realist's independently existing natural classes, or for a subjectivist's unique and universal set of artificially designed class constructs—this monograph shows that they have reached a dead end. The search has been a will o' the wisp, a search for something that not only does not exist, but cannot.

. . .

# 7. Appendix: A Framework-relative Recognition of a New Human Species

> *It may be dangerous to say this, for it could lead to indifferentism masquerading as broad-mindedness; where there is no one right answer, any answer may be regarded as more or less right.*
> – Phillips A. Griffiths (1958), p. 124

## 7.1 *Human Heterogeneity*

As we have seen, it is the specification of criteria of importance that determines whether commonalities may be recognized that might serve to define the classes of a taxonomy. Among these criteria, genetic resemblance has played a prominent role in many definitions of species. However, as we should by now expect from the logic underlying species definition, alternative criteria have been proposed, among them, and not without debate, *behavioral commonalities*. Peter and Rosemary Grant have made extensive studies of Darwin's ground finches of the Galápagos Islands. They have observed distinct breeding behaviors and songs among the birds, as well as morphological differences, and on this basis have claimed that the Galápagos ground finches comprise thirteen different species.[24] Also pointing to differentiation of species in terms of behavioral differences is the possibility of so-called heteropatric speciation, a subordinate form of sympatric speciation that, by some biologists, is believed may occur when organisms live in the same geographical area. Heteropatric speciation recognizes that behavioral differences among organisms that inhabit different ecological niches may lead to the development of new species (cf., e.g., Smith 1966, Getz & Kaitala 1989); in this sense, behavioral differences rather than geographic isolation provides the basis for speciation. Other biologists disagree—in connection with the alleged multiplicity of Galápagos finch species and/or heteropatric speciation—as a result, as one would also expect, of their commitments to alternative sets of criteria of importance and to corresponding evidence which they believe supports their choice of criteria.

---

[24] See, e.g.:
http://www-dept-edit.princeton.edu/eeb/people/display_person.xml?netid=prgrant, accessed 8/31/2015; Grant & Grant (1998, 2008); Ratcliffe & Grant (1983a, 1983b, 1985).

In this appendix, I apply a certain set of behavioral criteria to *Homo sapiens*, which today is believed to form but a single and homogeneous species. What follows is perhaps a radical thought experiment to illustrate, albeit in caricature, an implementation of the conclusions of this monograph. The application made here is intended in part to be taken seriously, but in part also as an exercise in intellectual fantasy that may be instructive. It is decidedly not proposed as a "test case" by which the logically supported conclusions we have reached stand or fall, since to interpret any one example in this way involves fallacious backwards reasoning (an individual example, suffering perhaps from its own inappropriateness or deficient portrayal, cannot undermine a rigorous general proof). Nevertheless, an illustration of this sort, purely because it may be regarded as extreme in its caricature—outrageous to some, but perhaps comforting to others—places in bold relief the logical exigencies the body of this monograph has sought to bring to light.

Before proceeding, it is intellectually bracing to keep in mind as a kind of stage backdrop the existing multiplicity of *kinds* of species definitions—ranging from the often-chosen biological species concept, to the cladistic species concept, the ecological, the evolutionary, the genetic, phylogenetic, the reproductive species concept, and their many varied bedfellows. At the same time, we might also have in mind the similarly long list of sobriquets that have been proposed for the present human species, sometimes in earnest, sometimes in jest or sarcasm; they include *Homo creator* ("creator man," suggested by Nicolaus Cusanus), *Homo demens* ("demented man," Edgard Morin), *Homo faber* ("man the toolmaker," Benjamin Franklin and Marx), *Homo ludens* ("playing man," Schiller), *Homo stultus* ("stupid man," Charles Richet), *Homo mendax* ("lying man," novelist Fernando Vallejo), *Homo sanguinis* ("bloody man," anatomist W. M. Cobb), and there are many more. —In the context of the pluralism of species definitions and the multiplicity of the human species' aliases, the christening of a new human species definition may seem less astonishing.

Against this backdrop it should come as no surprise that, in past millennia, some authors have occasionally contemplated the possibility, and some have argued for the reality, of a higher, evolutionarily more advanced human type that arises, now and again among us, in the form of exceptional men and women. From an evolutionary perspective, it is not inconceivable that more highly evolved human individuals might sporadically and, over long periods of time, gradually appear.

Of the many forms of speciation that have so far been stipulatively defined, the variety I would like to focus on here concerns what could be called *behavioral-dispositional speciation*. The term 'behavioral' is intended to cover not merely human overt physical behavior, but also cognitive behavior as it is manifested in cultural expression. By 'dispositional' I mean the likelihood, the propensity, that individuals and the groups to which they belong will make certain kinds of physical and cultural choices to behave in predictable ways. And, more specifically, I focus on the variety of human speciation that can occur within populations that live in close proximity or contact with one another—in other words, speciation that conceivably is now occurring among today's human populations.

When separate groups live in geographical separation, called allopatry, it has been widely accepted that new species are free, given adequate amounts of time, to evolve. But taxonomists have long argued whether speciation can occur between groups that live adjacent or in the same land area, correspondingly called parapatry and sympatry. For example, based on his observations in South America, and reinforced by his view that evolution occurs gradually, Darwin became convinced that new species can arise even though population groups live adjacent to one another. Since then, population genetics has supported Darwin's view (Turelli, Barton, & Coyne 2001, p. 337). And arising in large part from John Maynard Smith's work (Smith 1966), some biologists have also come to view sympatric speciation as likely.

The differentiable human groups that will be at issue in what follows most frequently live alongside one another, usually inhabiting different cultural-educational-economic-social "niches." Despite sharing the same geographical areas, they generally tend not to interbreed, but instead frequently engage in assortative reproduction—like individuals generally choosing others like them, that is, others who share commonalities perceived to be important.[25]

To turn back the clock for a moment: Plato, writing in his *Republic* more than 400 years before Christ, describes an ideal society consisting of three main *classes* of people (perhaps with a nod to taxonomy): producers, soldiers, and rulers. For Plato, the highest human attainment comes with the cultivation of the rational faculties, which are most developed in the class of rulers, and most fully realized in the philosopher.

Two millennia later, the basic idea continued to endure: Nietzsche is known for pointing to the more highly developed *Übermensch* as perhaps—he is vague—constituting a recognizably separate human species.[26] At one point, he remarked, "Schopenhauer wonders why Nature did not take it into her head to invent two entirely separate species of men" (Nietzsche 1911, Vol. 8, p. 159). It has been thought that Nietzsche's motivation in suggesting the possibility of two separate human species may have derived from his wish to emphasize the chasm separating the barbarians from the Greeks, and geniuses from the mediocre. That chasm may have inspired him to recognize the existence of the fundamentally different classes to which he perceived these people to belong.[27]

A similar thread reappears every so often, especially in philosophy, literature, and social criticism. In his 1895 novel, *The Time Machine*, H. G. Wells divided the future human population into two separate species, the Eloi and the Morlocks, who vary greatly in their behavior and values, in their "behavioral-dispositional characteristics": one group lives an enlightened, refined intellectual life aboveground in the open air and sun; the other, rude, savage, sordid, and filthy, leads a troglodytic existence in the underground dark.

Today, a century after Wells and reminiscent of his forecasted separation of human species, evolutionary theorist Oliver Curry of the London School of

---

[25] Cf., for example, the data relating to human assortative mating compiled by Herrnstein & Murray (1994).

[26] Whether Nietzsche intended the *Übermensch* to comprise a distinct human species is far from clear thanks to his often imprecise writing. It would require detailed textual analysis (perhaps) to answer this, but an answer is not relevant to the discussion here. Some Nietzsche scholars accept the distinct species interpretation; others, for example Kuderowicz (1976), do not.

[27] Golomb (2002), p. 67.

Economics similarly predicts that *Homo sapiens* will eventually split into two basically different species, an intellectually developed elite and a low class of the mediocre and dull-witted.[28] Paralleling this view, in the comparatively new area of research of so-called "human enhancement," already there are those who foresee a "post-human species": "a natural or artificially engineered lineage" with its own special characteristics (cf., e.g., Proust 2011, p. 150).

There clearly are precedents for the notion that *Homo sapiens* may be left behind; but none that fits the classical ideal that has come down to us from Plato quite as well as does a certain special group of commonalities that we shall describe.

## 7.2 *The Possibility of a New Human Species among Us*

We've noted the existence of a pluralism of species definitions and of alternatively conceived and named human species, remarked on the possibility of behavioral-dispositional speciation, and mentioned the views of Plato, Nietzsche, H. G. Wells, Oliver Curry, and proponents of human enhancement—this, then, is the fictional setting for the illustration I wish to hypothesize. In such a context we might be led to ask: Is the existence today of two fundamentally distinct, contemporaneous human species implausible and completely fanciful? Is there evidence that exemplars of "post-*Homo-sapiens*" occasionally arise among us?

As this monograph has shown, recognizing a set of commonalities that could be used to define such a new human species must of necessity themselves reflect a logically prior set of criteria of importance. What, in this speculation, might those criteria be? Is there sufficient and persuasive evidence to discriminate among the "objects in the domain" (that is, present-day people) so as to differentiate them into clearly defined separate classes—that is, into separate species? Are these criteria of identification justifiably important, and if so, to whom? What values do these criteria express? And, in the end, what purposes would the recognition of such a new human species among us accomplish?

Particularly in the history of literature, poetry, and art we find recurring expressions by many exceptional creative people of a sense of alienation from majority society. Often they are original thinkers whose intellectual or aesthetic efforts have placed them at odds with the world of ordinary people. These extra-ordinary individuals frequently experience a deep divide between themselves and others, between the kind of people they know themselves to be and their perception of the rest of humanity, between their values and interests and those of the ordinary population, between the propensities of others and their own dispositions to engage in certain kinds of activities, and to shun other activities that attract widespread and often passionate popular enthusiasm.

They have often been outcasts from "normal society," subject to its disapproval and censure and the suppression of their thought and work: We think of Socrates, Bruno, Copernicus, Savonarola, Galileo, and the many others, both major contributors and minor players, who throughout the history of civilization have suffered from the short-sightedness and ideological prejudice of their societies, their religious groups, their scientific peers, or their fellow academics.[29] And then there are those who have lived lives in lonely isolation from the "rest of humanity," with a self-awareness that recognized their special, extra-ordinary dissimilarity from others: We think of Virginia Woolf ("A Room of One's Own"—who was perhaps not far from wishing for "a species of one's own"), of Jean Cocteau ("one feels shame at being a part of the human race"[30]), of Romanian philosopher E. M. Cioran ("as far as I am concerned, I resign from humanity" (Cioran 1992/1934, p. 43), who later wrote of "freeing oneself of the Species, that hideous and immemorial riffraff" (Cioran 1998/1973, p. 108), and who asked "what was the use of being prized in a world inhabited by madmen, a world mired in mania and stupidity? For whom was one to bother, and to what end?" (Cioran 1998/1973, p. 26).

---

[28] Cf. http://news.bbc.co.uk/2/hi/uk/6057734.stm accessed 6/25/2015.

H. L. Mencken, not a social critic to mince words, made this comparative observation under the heading of "Varieties of *Homo sapiens*":

> [T]here are minds which never get any further than a sort of insensate sweating, like that of a kidney.... Of one mind we may say with some confidence that it shows an extraordinary capacity for function and development—that its possessor, exposed to a suitable process of training, may be trusted to acquire the largest body of knowledge and the highest skill at ratiocination to which *Homo sapiens* is adapted. Of another we may say with the same confidence that its abilities are sharply limited—that no conceivable training can move it beyond a certain point. In other words, men differ inside their heads as they differ outside. There are men who are naturally intelligent and can learn, and there are men who are naturally stupid and cannot....
>
> Men are not alike, and very little can be learned about the mental processes of a congressman, an ice-wagon driver or a cinema actor by studying the mental processes of a genuinely superior man. The difference is not only qualitative; it is also, in important ways, quantitative. (Mencken 1927/1926, p. 16-17, 22)

There have been many voices of writers, poets, artists, philosophers, and others raised in complaint against the human species, often expressing misanthropic bitterness and alienation. Their complaints and criticisms are frequently and casually dismissed by the larger society as the condemnations of ineffectual intellectuals and aesthetes, social malcontents who think of themselves as different and special, and who complain and criticize their societies probably in emotional compensation for their failures to "adjust," "fit in," and "be *normal*." And from the perspective of the so-called normal population, this is no doubt often true. From the extra-ordinary individual's perspective, however, the situation is reversed, as he or she frequently feels that the personal and creative adversity and hardship that come from having no other choice but to live with a population of insensitive, unintelligent, mediocre, and aesthetically deaf, dumb, and blind people is an imprisoning torture.[31] —This divide may sound a little bit like interspecies conflict....

---

[29] For a detailed discussion, see Bartlett (2011, Chap. 7).

[30] Jean Cocteau, in a 1963 interview by William Fifield published in 1964 in *Paris Review 32*, available at http://www.theparisreview.org/interviews/4485/the-art-of-fiction-no-34-jean-cocteau, accessed 6/10/2015.

The predictable appearance in the last paragraph of the word 'normal' should give us pause, for *if* occasional exemplars of post-*Homo-sapiens* do in fact arise among us, they no doubt would be labeled "abnormal" by the general, "normal" population. In contrasting the standards in terms of which we judge "normality" with those we use to identify "abnormality," we may find indications of the criteria of importance that are here in question. Since these standards apply largely to human characteristics that are psychological in nature, we need to shift attention to the concepts of psychological normality and abnormality.

In other publications,[32] I have examined the range of characteristics and dispositions in terms of which human psychological normality is understood by psychiatrists, psychologists, ethologists, and anthropologists. The conclusions reached include the following:

- psychological normality, far from serving as a gold standard for good psychological health, comprises a mixture of behavioral and dispositional characteristics that bring about much avoidable suffering and destruction of life;
- psychological abnormality, most often associated with mental illness, is a classification often applied not only vaguely, but wrongly when individuals are psychologically healthy, some of whom possess extraordinary capacities and creative abilities;
- there is a need for a clearer set of standards based on evidence in terms of which to judge both good and poor psychological health;
- statistically speaking, there are, as we shall note later, comparatively few people within the general population who can be regarded as having optimally good psychological health.

From the earlier discussion in this monograph, the reader should now be aware that recognition of those commonalities that are capable of defining a taxonomic species must necessarily rely upon judgments which have an extra-logical source and which reflect criteria that identify properties considered to be salient and important. A relatively small group of authors[33] has sought to direct attention to those defining properties which, from their observations, are descriptive of optimally positive human psychological health. A list of such properties can be used to formulate criteria of importance to identify and differentiate people in potentially distinct taxonomic classes.

---

[31] This perspective is examined in some detail in Bartlett (2009; 2011, Chap. 3, "The Abnormal Psychology of Creativity and the Pathology of Normality"; 2013a; 2013b).

[32] Bartlett (2005, 2011, 2013a, 2013b).

[33] For a detailed discussion of these authors' conceptions of positive psychological health, see individual chapters and sections in Bartlett (2005, 2011).

The following table summarizes a representative sampling of the principal properties which these authors have identified. The list is limited to authors who have made or implied a fundamental division between the small class of people who embody positive properties that are linked to good psychological health, as distinguished from the class of the majority of people.

| Author | Positive properties linked to good psychological health | Terms for those who are more advanced or psychologically healthy |
|---|---|---|
| Plato | <ul><li>cultivation of rational faculties</li><li>reflective critical thinking ability</li><li>independence of judgment</li></ul> | "the philosopher" (Cornford 1945, and other dialogues of Plato) |
| Schopenhauer | <ul><li>high level of individual moral development</li><li>conduct directed by compassion</li><li>exemplary attitudes and behavior</li></ul> | "individuals of superior insight" "people of moral worth" (Schopenhauer 1965/1841) |
| Nietzsche | <ul><li>opposition to mediocrity</li><li>independence of mind and judgment</li><li>moral autonomy</li><li>genuinely superior abilities, especially intellectual</li></ul> | "the *Übermensch*" (Nietzsche 1995/1885) |
| Lorenz | <ul><li>high intelligence</li><li>compassion</li><li>high moral worth</li></ul> | "the fully superior person" "the fully valuable person" (Lorenz 1940) |
| Fromm | <ul><li>individual autonomy</li><li>resistance to the pressures of conformity and enculturation</li></ul> | "the autonomous person" (Fromm 1941, 1947, 1955) |
| Jung | <ul><li>independent thought</li><li>autonomous moral judgment</li><li>resistance to conformity</li></ul> | "the individuated person" (Jung 1958) |

| Author | Positive properties linked to good psychological health | Terms for those who are more advanced or psychologically healthy |
|---|---|---|
| Maslow | <ul><li>individual autonomy</li><li>resistance to the pressures of conformity and enculturation</li><li>a sense of mission in life</li><li>responsive to beauty</li></ul> | "the superior person" (Maslow 1964)<br>"self-actualizing" (Maslow 1971)<br>"the more highly evolved person" (Maslow 1971) |
| Bartlett | <ul><li>individual autonomy, resistance to conformity, critical thinking ability</li><li>rationality wedded to conviction and consequent behavior</li><li>aesthetic sensibility</li><li>conviction that bridges moral reasoning and commitment to act accordingly</li><li>aversion to human aggression and destructiveness</li><li>fully functioning empathy toward humans and animals</li><li>avoidance of mere belief; elimination of conceptual pathology</li></ul> | "psychologically healthy, morally intelligent individuals" (Bartlett 2005, Chap. 18; 2011)<br><br>(Bartlett 2002)<br><br>"de-projective thinking" (Bartlett 1971; 1982; 1983; 2005, Chap. 3; 2011, pp. 45ff) |

**TABLE 2**

My purpose in bringing together the list of positive properties that appears in the middle column of the above table is to weigh the possibility of using these properties as criteria of importance by means of which to recognize a distinguishable class of men and women.

The reader will notice that several of the authors listed in the table have identified the same positive properties—for example, independent thinking, personal and moral autonomy, resistance to conformity. Readers who are skeptical of informal, semi-anecdotal observations made by any individual, however well-qualified, may be willing to grant that there is a measure of reliability to be found in the mutually confirming studies of these men, and others who are sprinkled among the millennia, who have invested large portions of their lives in an effort to understand what makes "a better human being."

More empirically oriented than most, Abraham Maslow sought to be very specific about the positive characteristics that he observed in "superior, self-actualizing" individuals. It is important to note that he showed that these positive properties can be defined operationally and measured by means of psychological testing, using, for example, the Personal Orientation Inventory (Shostrom 1963).

## 7.3 *A Few Questions and Objections*

Persuading the skeptical reader to accept the human importance of the listed positive characteristics is a matter of salesmanship that lies beyond the scope of this essay.[34] Instead, I turn to consider a group of interrelated questions: First, is it possible—theoretically as well as realistically plausible—to utilize criteria of identification based on the designated group of positive properties in order to differentiate and classify notably different kinds of people, and how would such classification be significant? Second, what should we expect to learn by doing this? Third, is such a classification useful? And last, is it valid? Let us take these questions one at a time.

Is it possible, and if so, how would such a classification be significant? Readers will see immediately that the positive characteristics listed in Table 2 pose evident short-comings: The majority of the characteristics are not easily evidenced or measured in a public way. Recognizing such characteristics is quite a different matter from being able to measure the size and color of the beaks of Galápagos ground finches, or to observe commonalities in reproductive behavior. These are openly visible characteristics; they do not require, for example, an individual psychiatrist's diagnostic judgment, or the results of a psychometric test, or an individual person's ability to attain accurate self-knowledge, in order to verify their presence in a person or a group.

Is our awareness of such properties therefore of little significance? There are several possible answers, none entirely satisfactory. In an important sense, a listing of these positive human characteristics can help to formulate an ideal, and by referring to that ideal to set goals more wisely, for example, in child-bearing, parenting, education, and in the social recognition of individuals who show these characteristics. Ideals are indispensable to purposeful, practical efforts.

A second way in which an awareness of the properties associated with positive psychological health may be significant concerns the personal self-awareness, validation and valuation of self, and self-understanding, which are essentially private matters that take place within the consciousness of individual persons. Recognizing the importance of a set of positive human characteristics such as Table 2's authors have sought to identify can contribute to a personal framework of meaning for exceptional individuals whose sense of identity and direction in life are admittedly often made more difficult by normal society.

Third, there is the statistical significance that comes from studies that have attempted to determine the incidence in the general population of individuals who have many of the characteristics associated with positive mental health. In Bartlett (2011, Appendix III, "The Distribution of Mental Health"), the observational results of a number of researchers have been brought together in chart form. These results come from studies of the incidence of both good and of poor mental health.

---

[34] Readers open to such persuasion may be interested in Bartlett (2005, 2011).

The outcome of plotting both sets of observations, and of applying positive criteria such as are listed in Table 2, leads to a recognition that only 3 to 6 percent of the present human population possesses genuinely positive levels of mental health (Bartlett, 2011, p. 276).

In connection with Maslow's studies of "superior, self-actualizing" people, he made the following comments, which have special relevance in the context of the discussion here: About these "remarkable human beings," he wrote: "*It was as if they were not quite people but something more than people*" (Maslow 1971, p. 42, italics added). During the many years of his work with such individuals, he confirmed their abovementioned statistical rarity, claiming that these individuals appear within the general population with only extreme infrequency, making up, in Maslow's count, "the healthiest 1 per cent or fraction of 1 percent" (Maslow, 1971, p. 92).[35] Maslow recognized that the very small size of this class of people would lead to a reaction on the part of his critics who will feel "deep conflicts over the 'elitism' that is inherent.... [T]hey are after all superior people whenever comparisons are made" (Maslow 1971, p. 289). In an earlier paper, he explained more clearly what concerned him:

> None ... dares to lock horns with the problem which is so unpopular in any democracy: that some people are superior to others, in any specific skill or—what is more provocative to the democrat—in *general* capacity. There is evidence that some people tend to be generally superior, that they are simply superior biological organisms born into the world. (Maslow 1964, p. 10)

I have touched on three possible answers to the first question, whether it is possible, using the specified group of positive properties, to differentiate a class of people who might be recognized as representatives or exemplars of a new, more developed human species, and if so what meaning such a classification might have. Yes, it is possible to distinguish such individuals from the general human population, but, at least at present, on an individual basis, this can be done only by means of the individual diagnostic testing and judgment of mental health professionals, or else by means of a person's own, often more fallible, self-understanding and self-evaluation. On a statistical basis, it is also possible to reach an estimate of the prevalence of what we may, *if* we choose, based on the combined observational evidence mentioned, refer to as "post-*Homo-sapiens*."

We turn to the remaining pair of questions, What should a taxonomic partitioning of the present human population lead us to recognize, and Is this recognition useful?

---

[35] Maslow's estimate of the uncommonness of such extra-ordinary people is complemented by the carefully researched conclusions reached by Charles Murray's systematic, comprehensive study of human achievement. In his book, Murray studied the incidence of genuine human eminence, which he also found occurs with extreme rarely. In his words: "When we assemble the human résumé, only a few thousand people stand apart from the rest. Among them, the people who are indispensable to the story of human accomplishment number in the hundreds. Among those hundreds, a handful stand conspicuously above everyone else" (Murray, 2003, p. 87).

As we have seen, the fundamental purpose of any taxonomy is to differentiate fundamental classes of organisms based on criteria of importance. In the case of human organisms, a taxonomic partitioning of the present human population leads one to recognize and therefore to distinguish individuals and groups of like individuals on the basis of their common possession, or lack of possession, of such positive characteristics as have been identified and selected for this purpose. We accept that such a selection cannot be derived purely logically, but reflects a *valuation and selection* of what is to be considered important. Here, of course, there exists complete freedom—in other words, inescapable ambiguity—to choose otherwise.

However, *if* the positive characteristics summarized in Table 2 are accepted as criteria of commonality recognition, *then* it will follow that people can be classified accordingly, thereby recognizing them as comprising a distinct group. This is the self-validating result that we should expect once we have chosen to recognize these particular characteristics as important.

This fact expresses nothing more than a tautology, and cannot *not* be accepted.[36] We do not, however, mistake this fact to imply that the hypothesized classification *must* in any way be accepted. Its acceptance simply cannot be avoided *provided* the criteria of identification are applied. This is no logical sleight-of-hand; rather, it expresses in a taxonomic framework the relationship that, in a parallel fashion, also exists between any formal system's axioms and the results derivable from them. Once the formal framework in which those axioms are expressed is accepted, the derived results cannot *not* be accepted. As a result, we have in mathematics distinct and alternative systems of geometry, much as we do different taxonomic classification systems. The useful applicability of these systems to real world problems is of course a different matter.

The usefulness of a particular taxonomic classification is determined by the degree that it satisfies a set of values. Previous mention has been made of the valuation component of pattern recognition; we need to return briefly to that subject. As we have seen, whether one values the positive characteristics at issue here has no purely logical justification. In part this is because the identification of those characteristics as significant and important already has built into it an evaluative decision. To insist on an answer to the question, On the basis of what values should these positive characteristics be valued?, is to demand an answer outside a vicious circle while requiring that one remain in it. In this sense, the set of positive characteristics in question is no longer subject to question as soon as these characteristics are identified as "positive" and therefore meaningful and valuable.

Finally, to turn to the question whether the classification "post-*Homo-sapiens*" is a valid classification: The framework-relative approach which this monograph acknowledges leads directly to an affirmative answer. The present appendix itself provides a reference frame in terms of which a group of properties is specified, which we are then free to employ as criteria of identification; their application makes it possible to discriminate among individuals and groups of like individuals on the basis of a recognition of commonalities judged to be important. This is what the validity of any classification ultimately means.

---

[36] See footnote 23.

Shall we, then, call the resulting differentiable classes of people different "species"?

There are various foreseeable objections to the drawing of species dividing lines within a single, apparently homogenous, freely interbreeding population such as the human one today. One of the most famous attempts to weigh both sides of this question was Darwin's in Chapter VII of *The Descent of Man* in a section titled "On the Races of Man." Tracing the course of Darwin's erratic and somewhat bewildering thinking can be instructive from a perspective a century and a half later.

In that section of his book, Darwin confronted the question whether the various human races should be classified as different species. He wrote that it was his intention "to inquire what is *the value of the differences between them under a classificatory point of view*, and how they have originated" (Darwin 1871, p. 214, italics added).

Darwin's reasoning concerning racial differences is only marginally relevant to the discussion here, which, far from having different human races in view as did Darwin, concerns rather a group of special, positive human behavioral dispositions *that cut across racial distinctions*. His thinking process, nevertheless, is of interest and can be informative.

While the main concern in this appendix has been to consider the propensities of individuals and their groups that lead them to make certain kinds of physical and cultural choices, these propensities are not unique to any race, nor to present knowledge do they correlate with racial distinctions. Darwin's observations concerned racial differences rather than the differences that have drawn our attention in this appendix; the two types of differences define fundamentally different areas of study.

With this caveat, let us look more closely at Darwin's reasoning. In his usual, patient, and cautious manner, he began by identifying a number of key ways in which human races differ—e.g., in hair texture, lung capacity, physical constitution, vulnerability to certain diseases, geographical distribution, interbreeding, even in having their own distinctive external parasites (Darwin 1971, pp. 216ff). More relevant to our focus here is his observation, "Their mental characteristics are likely very distinct; chiefly as it would appear in their emotional, but partly in their intellectual, faculties" (p. 216). Here, Darwin began to single out, to select, properties relating to the category of mental characteristics, and, as we'll see shortly, he decided to favor what I have called behavioral-dispositional characteristics as identification criteria in order to determine human taxa.

Having identified a variety of ways in which the human races differ, on this basis he recognized "that a naturalist might feel himself fully justified in ranking the races of man as distinct species; for he has found that they are distinguished by many differences in structure and constitution, some being of importance" (p. 224). However, he then quickly began to wonder whether it might be more appropriate to regard races as "sub-species": "if we reflect on the weighty arguments ... for raising the races of man to the dignity of species, and the insuperable difficulties on the other side in defining them, the term 'sub-species' might here be used with much propriety. But from long habit the term 'race' will perhaps always be employed" (p. 228).

After this suggestion to allow the question to dissolve into established habit, he made the following comment, quoted earlier in the main text of this monograph: In attempting to answer the question whether the human population should be classified into a plurality of species (or sub-species) according to race, Darwin acknowledged that:

> ... it is a hopeless endeavour to decide this point on sound grounds, until some definition of the term 'species' is generally accepted; and the definition must not include an element which cannot possibly be ascertained, such as an act of creation. We might as well attempt without any definition to decide whether a certain number of houses should be called a village, or town, or city. (p. 228)

But then, having first proposed that we continue to use the term 'race' rather than 'species' or 'sub-species', and then deciding that it is a "hopeless endeavour" to decide the issue, he then went on to provide a detailed argument that seeks to emphasize human homogeneity. To do this he cited *similarities* "between the men of all races in tastes, dispositions, and habits" (p. 232). These similarities include dancing, music, acting, painting, tattooing, use of gestures, physical and vocal expression. (—N.B. All of these are what I have termed behavioral-dispositional characteristics.) Such similarities that are common to the various races should, Darwin suggested, lead us to recognize the homogeneity of the human species:

> ... when naturalists observe a close agreement in numerous small details of habits, tastes and dispositions between two or more domestic races, or between nearly-allied natural forms, they use this fact as an argument that all are descended from a common progenitor who was thus endowed; and consequently that all should be classed under the same species. The same argument may be applied with much force to the races of man." (p. 233)

This passage gives the appearance that Darwin has reached a final conclusion, but he again turns out to be ambivalent and wavers. Two pages later, he threw in the towel, and wrote: "it is almost a matter of *indifference* whether the so-called races of man are thus designated, or are ranked as species or sub-species; but the latter term appears the most appropriate" (p. 235, italics added). It looks almost as if Darwin embraced the thought expressed by Phillips A Griffiths (1958, p. 124) quoted at the beginning of this appendix: "It may be dangerous to say this, for it could lead to indifferentism masquerading as broad-mindedness; where there is no one right answer, any answer may be regarded as more or less right."

What can we learn from Darwin's example? Here is how I understand Darwin's wavering process of thinking and what I draw from it:

(i) It seems clear that Darwin did not *want to believe* in human heterogeneity despite his acknowledgment of the various ways in which racial groups differ. (ii) If pressed, in light of evident differences among races, he was willing to apply the term 'species', or preferably 'sub-species', but was content simply to continue to use the habitual word 'race'—in other words, the choice of word, from his perspective, as we might say today, appeared to him to be a matter of semantical preference and convenience. (iii) *But* he would prefer to recognize and emphasize the many ways in which racial groups are *similar*, and here he pointed to commonalities of tastes, dispositions, and habits. (iv) These commonalities are, in his judgment, important, and due to their importance, he chose to give the greater weight to these particular shared characteristics rather than the variety of characteristics that differentiate the races. Based on this weighting, he therefore recommended that the different human races be considered members of one unified human species.

We should note that this course of thinking, as I have described it, is filled with evaluative terms that express implicit value judgments, something that we have learned is at the core of the pattern recognition of commonalities. Also prominent is the characteristically human tendency to *want to believe* certain things that reflect and express fundamental value commitments. I do not make these comments in criticism, but to point out that they are to be expected whenever judgments of commonalities are made.

If we return to the theme of this appendix, Darwin's example is instructive: He, too, observed that different human groups often exhibit distinct "mental characteristics," primarily emotional, but in part intellectual. It was characteristics of this sort that he clearly had in mind when his thinking reached (iii) and (iv) above. The human sharing of common tastes, dispositions, and habits came to have a decisive importance for Darwin in claiming that racial differences should be subordinated to them.

This parallels in a significant way the priority given in this appendix to human "behavioral dispositions," a category which includes "tastes, dispositions, and habits." But rather than finding that the positive characteristics summarized in Table 2 *unify* human populations, as Darwin would probably have preferred to believe, these characteristics, as we have seen, instead point to fundamental distinctions that differentiate people.

In considering the question whether there are multiple human species, Darwin's course of thinking proceeded haltingly and ambivalently, reflecting, as we now know, the inescapability of ambiguity that affects all judgments of commonality. A close look at his handling of the question brings out that *any* choice of criteria of identification, which must be relied upon in order to differentiate distinct species, rests on a prior affirmation of a set of values. Somewhat paradoxically, the high valuation that Darwin gave to human behavioral-dispositional characteristics pointed for him to human homogeneity and a single human species, whereas the choice—if we make it—to value Table 2's set of positive dispositional characteristics has pointed in the very opposite direction, to human heterogeneity and perhaps to the plausibility of ongoing speciation.

Whether to accept human behavioral-dispositional speciation is a question that can only be answered, then, in terms of the values one endorses and which determine one's criteria of identification. Some taxonomists will prefer their own framework-relative commitments that do not accept Table 2's set of identification criteria. I attempt to give no response to those who take this position because, in light of the preceding discussion, none can be given that is not pointless or does not beg the question: A response to an opposing taxonomist's framework-relative commitments must either accept that opposing framework (which is guaranteed to miss the point this monograph seeks to make), or it does not accept that framework (which would beg the taxonomist's question).

A foreseeable possible objection to ongoing human speciation stems from the thought that if there presently exist two differentiable human species, this could prevent their interbreeding, but interbreeding, as we know, is something all human populations engage in. For readers not laden with the baggage of conflicting prior commitments and who may be unaware of some of the relevant biology, they are reminded of the following well-established fact: Some classes of organisms that are considered to be distinct species can and do interbreed (for example, lions and tigers do, primarily when caged together; as well as dogs, jackals, and coyotes; cattle with bison; some of the Galápagos finch species with one another; *H. sapiens* with certain archaic hominins, e.g., with *H. neanderthalensis* (as presently thought); etc.); therefore there should in principle be no species-based obstacle to *Homo sapiens*/post-*Homo sapiens* reproduction.

Furthermore, classes of organisms may be considered to be distinct species despite having a common genetic blueprint and place in an evolutionary map of descent; the Grants' pluralism of Galápagos ground finch species may, for example, point in this direction. And as we also noted earlier in connection with parapatric and sympatric speciation, new species may arise despite the fact that population groups live in close proximity to one another; here again there is no species-based obstacle to ongoing human speciation. The fact that extra-ordinary individuals who possess the set of positive characteristics we have identified often select mates that have those characteristics in common lends further assortative support to the possible development among us of a differentiable, perhaps still relatively nascent, new species.[37]

Depending, then, upon the values one endorses, recognizing a new human species that possesses the positive characteristics we have discussed may come as good news. Darwin believed that natural selection is progressive, leading to resulting improvements in an organism—that is to say, if environmental conditions are stable, he believed that natural selection would increase the degree of adaptation of a population. The modern view of natural selection, however, has come to assert that natural selection does not necessarily result in organism/environment optimization (e.g., Sober 1985, p. 868). Whether a new human species, such as the one hypothesized here, will show an improvement in its environmental adaptation is clearly open to question; some critics of the alleged positive qualities listed in Table 2 would likely doubt this. Furthermore, even granting that exemplars of such a new human species may in fact occasionally stand among us, this is no guarantee that natural selection, possibly in the form of the continued historical, deliberate suppression of the new human species by the old, will operate to insure its survival.

---

[37] See footnote 24.

## 7.4 *Concluding Unscientific Postscript*

The result of this fanciful thought experiment is an informally expressed theory along with the description of a model, a domain of objects, which substantiates true statements made in the theory. The theory here is a conceptual expression of the observations made by Plato, Schopenhauer, and others in their efforts to identify what is best in humanity as this is embodied in extra-ordinary individuals. The application of the set of criteria based on their observations leads, if we wish, to an informative conceptual division of people into two distinguishable groups, a division that at present is often difficult to discern in real life on an individual basis. Whether we are willing to call these two groups different "species" has no logically compelling answer. If doing so contributes to one's desired purposes, then this makes sense; otherwise of course it doesn't. But then this is no different than it is with any other species definition.

What might those desired purposes be? We live at a time when there is social and political pressure to ignore or gloss over individual human differences, most especially the intellectual variety, and to espouse the belief that all people may—if they choose, and granting the proper environmental conditions—develop however they wish, all with an equal potential to excel. Long before cheers went up for the No Child Left Behind policy, as Leon Trotsky cheerfully postulated in his book *Literature and Revolution*, "the average human type will rise to the heights of an Aristotle, a Goethe, or a Marx" (quoted in Nozick 1974, p. 241).

Whether this buoyant belief is actually true lies beyond the scope of this monograph; however, we should recognize that, at the present time, any discussion of ongoing human speciation can, by some, be misinterpreted as an elitist challenge to a belief in the equal potential of all: i.e., to be whatever they wish if only external conditions are right. The evolutionary processes of speciation—of the branching development and proliferation of one class of organisms into a plurality of distinguishable species—are, however, not processes, so far as we know, that lend themselves to individual choices and initiatives, and so there is no challenge here to what an individual can or cannot accomplish in his or her life.

But given the current association of studies of individual human differences with political incorrectness, it may therefore at this time not be judged useful, or socially and politically expedient, to differentiate people within a taxonomic framework on the basis of their positive qualities. If we choose not to do this because we have, in a wholesale manner, rejected all research concerning individual differences, we may risk losing sight of an ideal that has compelled the respect and admiration of many throughout civilization's history. And if, furthermore, we choose not to do this, we also risk the loss by our educational system of an ideal that has been fundamental to the conception of "higher" in the phrase "higher education."[38]

On the other hand, if we do decide to distinguish levels of individual psychological health and attainment on the basis of such positive characteristics as have been identified here, and on this basis judge that a classification of people into two distinguishable human species reflects an important factual recognition of their differences, then we make a choice that is essentially and profoundly "discriminatory." For this would be a choice that logically leads to discriminatory consequences both in theory and in practice—and yet this choice and its consequences would not be "discriminatory" in its prevailing sense today.

---

[38] For a detailed study of these subjects, see Bartlett (1994a; 1994b; 2011, Chap. 5).

'Discrimination' is a word that has received bad press during the past several decades; the word now brings about a reflex-arc of emotionality and high blood pressure that cloud perception. At an earlier time, *to be discriminating* was to possess a well-articulated set of values and consequent good taste. That meaning of the word has generally fallen by the wayside in favor of its negative big brother that expresses reprehensible and baseless prejudice. Here, once again, we meet a variety of ambiguity.

And so it is—having reached non-ambiguous, definite, and provable conclusions regarding the inescapabilities of the ambiguities of species definition, of the pattern recognition of commonalities, of theories in relation to their models, and of matters of taste and bias—that this half-serious, half-whimsical thought experiment must draw to an end, and with it this essay as well.

...

# ADDITIONAL SELECTED SOURCES

# ABOUT THE AUTHOR

Homepage: http://www.willamette.edu/~sbartlet

Steven James Bartlett was born in Mexico City and educated in Mexico, the United States, and France. His undergraduate work was at the University of Santa Clara and at Raymond College, an Oxford-style honors college of the University of the Pacific. He received his master's degree from the University of California, Santa Barbara; his doctorate from the Université de Paris, where his research was directed by Paul Ricoeur; and has done post-doctoral study in psychology and psychotherapy at Saint Louis University and Washington University. He has been the recipient of many honors, awards, grants, scholarships, and fellowships. His research has been supported under contract or grant by the Alliance Française, the American Association for the Advancement of Science, the Center for the Study of Democratic Institutions, the Lilly Endowment, the Max-Planck-Gesellschaft, the National Science Foundation, the Rand Corporation, and others.

Bartlett brings to the present work an unusual background consisting of training in microbiology, pathology, psychology, and epistemology. He is the author or editor of sixteen books and monographs, and many papers and research studies in the fields of psychology, epistemology, mathematical logic, and philosophy of science. He has taught at Saint Louis University and the University of Florida, and has held research positions at the Max-Planck-Institute in Starnberg, Germany and at the Center for the Study of Democratic Institutions in Santa Barbara. He has received honorary faculty research appointments from Willamette University and Oregon State University, and now devotes full-time to research and publication.